\RequirePackage{lineno}

\documentclass[floatfix, noshowpacs, preprintnumbers, twocolumn, amsmath, amssymb, aps, prl, superscriptaddress]{revtex4}

\usepackage{graphicx, xcolor}
\usepackage[caption=false]{subfig}
\usepackage{soul}
\usepackage{graphics,amssymb,amsmath,epsfig,color}
\usepackage{mathrsfs}
\usepackage{titlesec}

\let\oldequation\equation
\let\oldendequation\endequation

\renewenvironment{equation}
  {\linenomathNonumbers\oldequation}
  {\oldendequation\endlinenomath}

\begin{document}


\title{Simulating Photosynthetic Energy Transport on a Photonic Network}

\author{Hao Tang}
\altaffiliation{These authors contributed equally to this work.}
\affiliation{Center for Integrated Quantum Information Technologies (IQIT), School of Physics and Astronomy and State Key Laboratory of Advanced Optical Communication Systems and Networks, Shanghai Jiao Tong University, Shanghai 200240, China}

\author{Xiao-Wen Shang}
\altaffiliation{These authors contributed equally to this work.}
\affiliation{Center for Integrated Quantum Information Technologies (IQIT), School of Physics and Astronomy and State Key Laboratory of Advanced Optical Communication Systems and Networks, Shanghai Jiao Tong University, Shanghai 200240, China}

\author{Zi-Yu Shi}
\affiliation{Center for Integrated Quantum Information Technologies (IQIT), School of Physics and Astronomy and State Key Laboratory of Advanced Optical Communication Systems and Networks, Shanghai Jiao Tong University, Shanghai 200240, China}

\author{Tian-Shen He}
\affiliation{Center for Integrated Quantum Information Technologies (IQIT), School of Physics and Astronomy and State Key Laboratory of Advanced Optical Communication Systems and Networks, Shanghai Jiao Tong University, Shanghai 200240, China}

\author{Zhen Feng}
\affiliation{Center for Integrated Quantum Information Technologies (IQIT), School of Physics and Astronomy and State Key Laboratory of Advanced Optical Communication Systems and Networks, Shanghai Jiao Tong University, Shanghai 200240, China}

\author{Tian-Yu Wang}
\affiliation{Center for Integrated Quantum Information Technologies (IQIT), School of Physics and Astronomy and State Key Laboratory of Advanced Optical Communication Systems and Networks, Shanghai Jiao Tong University, Shanghai 200240, China}

\author{Ruoxi Shi}
\affiliation{Center for Integrated Quantum Information Technologies (IQIT), School of Physics and Astronomy and State Key Laboratory of Advanced Optical Communication Systems and Networks, Shanghai Jiao Tong University, Shanghai 200240, China}

\author{Hui-Ming Wang}
\affiliation{Center for Integrated Quantum Information Technologies (IQIT), School of Physics and Astronomy and State Key Laboratory of Advanced Optical Communication Systems and Networks, Shanghai Jiao Tong University, Shanghai 200240, China}

\author{Xi Tan}
\affiliation{Center for Integrated Quantum Information Technologies (IQIT), School of Physics and Astronomy and State Key Laboratory of Advanced Optical Communication Systems and Networks, Shanghai Jiao Tong University, Shanghai 200240, China}

\author{Xiao-Yun Xu}
\affiliation{Center for Integrated Quantum Information Technologies (IQIT), School of Physics and Astronomy and State Key Laboratory of Advanced Optical Communication Systems and Networks, Shanghai Jiao Tong University, Shanghai 200240, China}

\author{Yao Wang}
\affiliation{Center for Integrated Quantum Information Technologies (IQIT), School of Physics and Astronomy and State Key Laboratory of Advanced Optical Communication Systems and Networks, Shanghai Jiao Tong University, Shanghai 200240, China}

\author{Jun Gao}
\affiliation{Center for Integrated Quantum Information Technologies (IQIT), School of Physics and Astronomy and State Key Laboratory of Advanced Optical Communication Systems and Networks, Shanghai Jiao Tong University, Shanghai 200240, China}

\author{M. S. Kim}
\affiliation{QOLS, Blackett Laboratory, Imperial College London, London SW7 2AZ,  UK}
\affiliation{Korea Institute of Advanced Study, Dongdaemoon-Gu, Seoul 02455, South Korea}

\author{Xian-Min Jin}
\email{xianmin.jin@sjtu.edu.cn}
\affiliation{Center for Integrated Quantum Information Technologies (IQIT), School of Physics and Astronomy and State Key Laboratory of Advanced Optical Communication Systems and Networks, Shanghai Jiao Tong University, Shanghai 200240, China}
\affiliation{TuringQ Co., Ltd., Shanghai 200240, China}


%

\maketitle

\textbf{
Quantum effects in photosynthetic energy transport in nature, especially for the typical Fenna-Matthews-Olson (FMO) complexes, are extensively studied in quantum biology. Such energy transport processes can be investigated as open quantum systems that blend the quantum coherence and environmental noises, and have been experimentally simulated on a few quantum devices. However, the existing experiments always lack a solid quantum simulation for the FMO energy transport due to their constraints to map a variety of issues in actual FMO complexes that have rich biological meanings. Here we successfully map the full coupling profile of the seven-site FMO structure by comprehensive characterization and precise control of the evanescent coupling of the three-dimensional waveguide array. By applying a stochastic dynamical modulation on each waveguide, we introduce the base site energy and the dephasing term in colored noises to faithfully simulate the power spectral density of the FMO complexes. We show our photonic model well interprets the issues including the reorganization energy, vibrational assistance, exciton transfer and energy localization. We further experimentally demonstrate the existence of an optimal transport efficiency at certain dephasing strength, providing a window to closely investigate environment-assisted quantum transport.
}

~\\
Since experimental evidences for quantum coherent energy transport have been successively observed in many pigment-protein complexes \cite{Engel2007, Lee2007, Collini2010, Panitch2010}, the photosynthetic light-harvesting process began to be investigated as open quantum systems that blend the quantum coherence and environment noises \cite{Breuer2007, Mohseni2008, Plenio2008, Caruso2009, Caruso2014, Wu2010}. The theory on  Environment-assisted quantum transport (ENAQT)\cite{Rebentrost2009} was then raised to suggest the enhancement of energy transport efficiencies by environment noises in many nanoscale transport systems. \textcolor[rgb]{0,0,0}{The ENAQT theory has been extensively studied and showed good interpretability on energy transfer among many coherent and incoherent theories\cite{Tao2020a,Park2016}.} The ENAQT theory has now been applied to a rich range of research areas including light-harvesting phenomena in nature, the solar cell engineering and other novel biotic excitonic devices\cite{Scholes2011}. 

One of the most well-studied natural creatures for its light-harvesting process is the green-sulphur bacteria, since its structure is simple but highly effective to allow for enough harvest of energy from the very dark deep sea environment\cite{Lambert2013}. The bacteria collect light through their large chlorosome antenna and transfer excitons to their reaction centre. The cable connecting these two part is the so-called Fenna-Matthews-Olson (FMO) complex\cite{Fenna1975}, which is normally formed in a trimer of three complexes with each complex consisting of eight bacteriochlorophyll a (BChl-a) molecules\cite{Wu2010}. Seven of the eight molecules are bound within a protein scaffold, which forms the environment for the complexes and provides the source of noise and decoherence. The eighth BChl outside the protein scaffold assists the transport of the excitation into the seven-site structure\cite{Hase2017}, where the seven BChls are conventionally numbered from No. 1 to 7 (Fig.1a). The seven-site FMO complex is a prevalent structure for the exciton transfer process. The excitation energy normally transports from BChl 1 or BChl 6 all the way to BChl 3, and eventually goes to the reaction centre to accomplish the energy conversion reactions for photosynthesis. 

\begin{figure*}[ht!]
\includegraphics[width=0.7\textwidth]{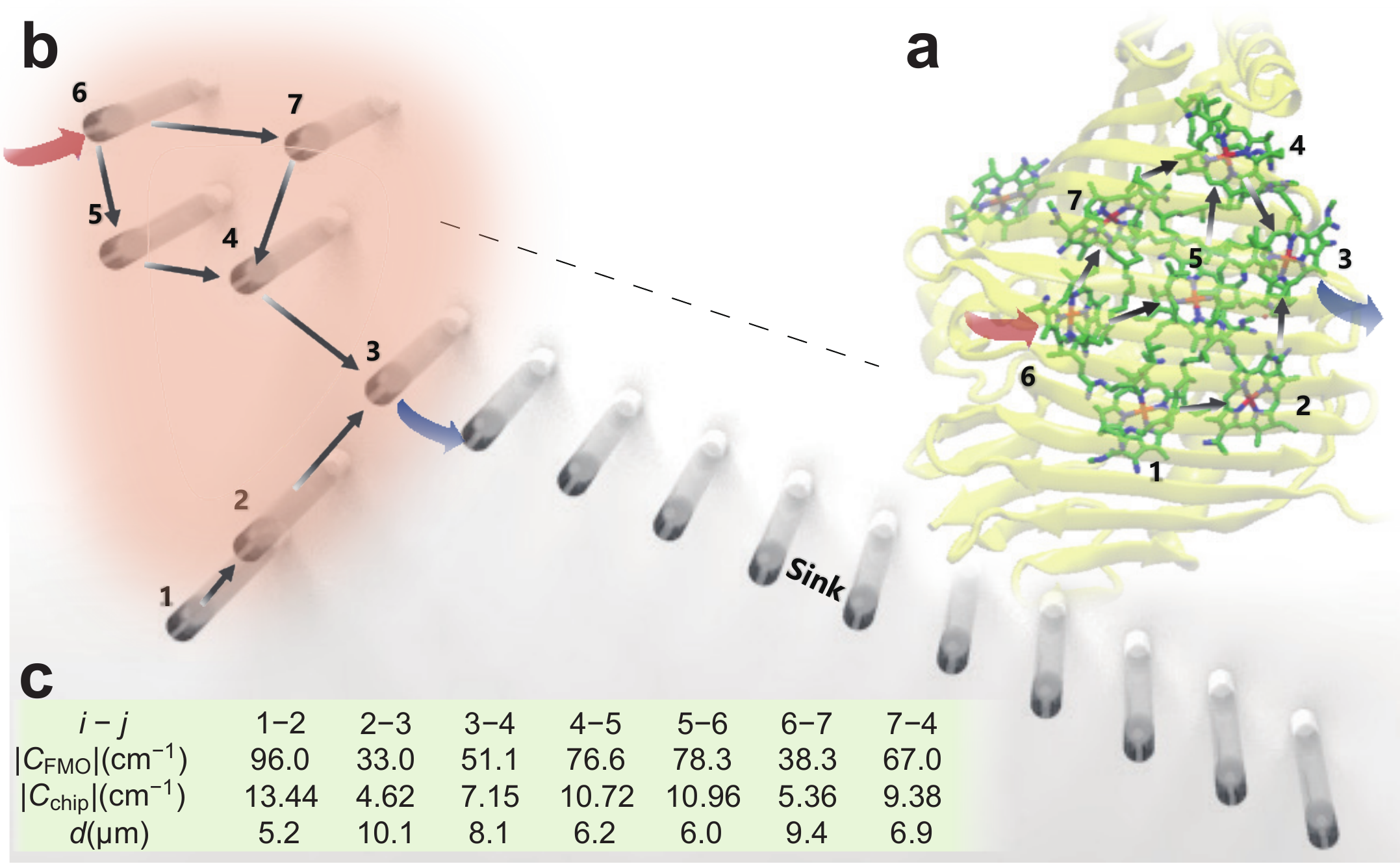}
\caption{\textbf{Experimental layout.} Schematic diagram of ({\bf a}) the FMO complex (Protein Data Bank accession 3ENI\cite{PDB}) and ({\bf b}) the three-dimensional photonic waveguide array simulating the FMO complex. The numbers of the 7 sites in the FMO complex and their corresponding waveguide are marked. The arrows show that the energy comes into the FMO complex or the waveguide array from Site 6 and moves through Site 3 to the sinks. ({\bf c}) The absolute values of the coupling coefficient in the FMO complex of $C. tepidum$, $C_{\rm {FMO}}$ (unit: cm$^{-1}$), according to \cite{Adolphs2006, Hoyer2010}. In order to map $C_{\rm {FMO}}$ onto the photonic lattice of a suitable propagation length, $C_{\rm {chip}}$ (unit: cm$^{-1}$), all the coupling coefficients on chip, are proportionally reduced to 14$\%$ of $C_{\rm {FMO}}$, which only affects the overall evolution time, but not the coupling profile among the seven sites. $d$ (unit: $\rm{\mu}$m) is the center-to-center waveguide spacing between two waveguides. Such $d$ values are set to generate the expected $C_{\rm {chip}}$ values above, as $C_{\rm {chip}}$ exponentially decays with $d$, which has been fitted by: $C=47.19\times e^{-0.2243d}$. The coupling coefficients between other sites are much weaker (below 15cm$^{-1}$), causing very marginal influences on the evolution pattern, and hence are not shown in the table. }
\label{fig:QFTConcept}
\end{figure*}

Theoretical quantum physicists have investigated the FMO structure \cite{Mohseni2008, Plenio2008, Caruso2009, Caruso2014, Wu2010} and proven that there exists certain optimal environmental noise levels to assist for an optimal energy transport efficiency. In recent years, many experimental simulations for ENQAT, especially in the context of a photosynthetic complex model, have emerged \cite{Biggerstaff2016, Harris2017, Potocnik2018, Wang2018, Tao2020b, Maier2019}. They are implemented in different systems, including a programmable nanophotonic processor with discrete-time evolution \cite{Harris2017}, superconducting circuits\cite{Potocnik2018}, the nuclear magnetic resonance \cite{Wang2018}, \textcolor[rgb]{0,0,0}{and the ion-trap qubits\cite{Maier2019}}, with a key goal on introducing controllable environmental noise into the original quantum system\cite{Tang2020}. However, there lacks a solid mapping to the FMO photosynthetic energy transport due to various constraints. Firstly, the quantum simulator hardware did not load the full Hamiltonian matrix for the authentic FMO in nature, since simulating the coupling profile for the seven BChls demands strong capabilities on setting the two-dimensional coupling space and flexibly tuning the coupling strength. Secondly, the issue on noises for FMO was not addressed. Some simply uses white noises\cite{Biggerstaff2016, Harris2017, Potocnik2018}, while some analyzes colored noises\cite{Wang2018, Maier2019}, which still does not match a spectral density for real FMO. Thirdly, there are many up-to-date works on photosynthetic energy transport\cite{Cao2020, Thyrhaug2018, Mancal2020}. Many important issues like the reorganization energy dynamics and the assistance by vibrational coherence \cite{Ishizaki2021, Wendling2000, Klinger2020, Ishizaki2009,Jang2008} are left for further investigations.

In this work, we present a close investigation on simulating photosynthetic energy transport in our three-dimensional photonic lattice. We, for the first time, map the full coupling profile of the seven-site FMO structure on a physical quantum simulator. Besides, by implementing the $\Delta \beta$ photonic model \cite{Caruso2016, Perez2018, Tang2019, Tang2022}, we introduce independently controllable noise for each waveguide, which allows us to load the site energy and the colored noise that yields a spectral density consistent with that for the actual FMO complex. We demonstrate the photonic model can simulate important biological issues including the reorganization energy and the vibrational assistance. Furthermore, by mapping the Hamiltonian matrix and colored noises for FMO on our photonic lattice, we carry out a quantum simulation experiment and demonstrate that an optimal energy transport efficiency exists at a certain noise amplitude. Our work provides a window to closely investigate environment-assisted quantum transport, and may inspire further explorations on comprehensive mechanisms for light-harvesting photosynthesis. 

Three-dimensional photonic array provides a highly versatile platform for quantum simulation. The longitudinal direction corresponds to the evolution time and the cross-section of the array structure could be engineered to implement a designed Hamiltonian matrix. \textcolor[rgb]{0,0,0}{Photons propagating through an array of $N$ coupled waveguides can be described by a $N\times N$ Hamiltonian matrix: 
\begin{equation}\label{H}
H= \sum_{i}^N \beta_{i}a_{i}^\dagger a_{i} +\sum_{j\neq i}^N C_{i,j} (a_{i}^\dagger a_{j}+a_{j}^\dagger a_{i}),
\end{equation}
where the diagonal values of $H$ are $\beta_i$s, the propagating constant along the $i$th waveguide,} and the off-diagonal terms are $C_{i,j}$, the coupling coefficient between waveguide $i$ and $j$.  

We use seven waveguides to represent the seven sites of the FMO complex (See Fig.1a and 1b). Since the coupling coefficients between two adjacent waveguides, $C_{\rm {chip}}$, are characterized to follow an exponential decay with the center-to-center waveguide spacing $d$ \cite{Tang2018, Tang2018b, Wang2022}, we are able to quantitatively control $C_{\rm {chip}}$ by carefully designing the waveguide configuration (See Fig.1c). Therefore, utilizing the two-dimensional evolution space, we faithfully map the major coupling coefficients for the real seven-site FMO complex of $C. tepidum$\cite{Adolphs2006, Hoyer2010} on chip. The full information on the Hamiltonian matrix for the FMO molecule is given in Supplementary Note 1. An extra array of 100 waveguides is connected to Waveguide 3 to serve as the sink, resembling the light transport from Site 3 of the FMO complex to its reaction center (Fig.1b). 

{\it The $\Delta \beta$ Photonic Approach.} Our photonic array is essentially a quantum evolution system for pure quantum walks, if all propagation constants $\beta_i$s in Eq.\ref{H} remain constant in time. Here we manage to modulate the diagonal term of the Hamiltonian by introducing $\Delta \beta$  , the detunings of the propagation constant $\beta$, in order to create a fluctuation of the site energy\cite{Caruso2016, Tang2019, Perez2018, Tang2022} (Fig.2a). A large number of stochastic $\Delta \beta$ detunings constitute the quantum stochastic walk\cite{Tang2022} and faithfully implement the dephasing process in the open quantum systems\cite{Rebentrost2009,Perez2018}. The introduction of $\Delta \beta$ can be experimentally achieved by tuning the laser writing speed during the waveguide fabrication process (see details in Supplementary Note 2). The existence of $\Delta \beta$ also causes some fluctuations of the effective coupling coefficent denoted as $\Delta C$, but the value is minor and it is shown to cause very marginal influence on the transport efficiency. We hence mainly consider the model with only diagonal $\Delta \beta$ terms (see discussions on $\Delta C$ in Supplementary Note 3). 

\begin{figure}[b]
\includegraphics[width=0.48\textwidth]{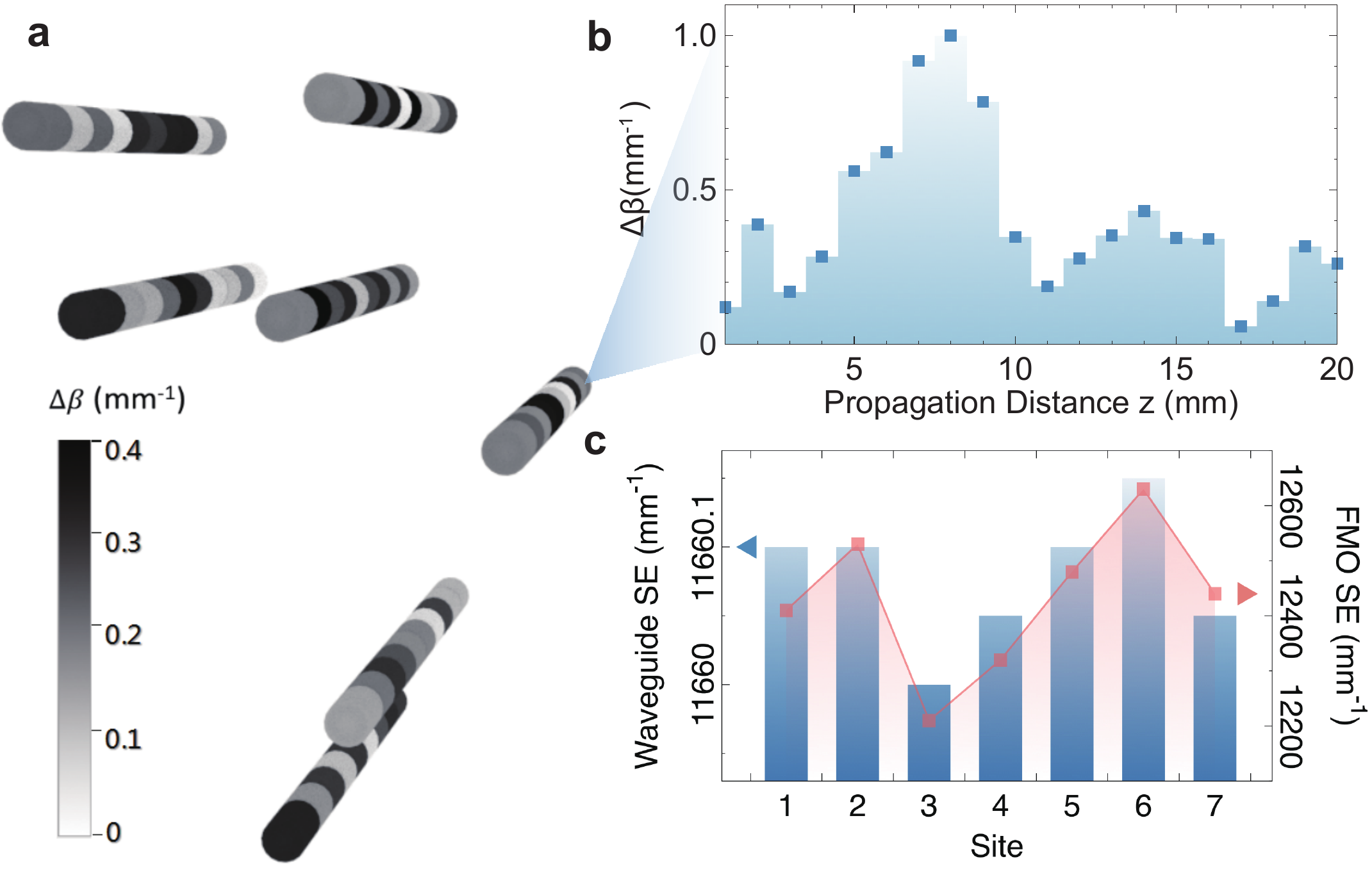}
\caption{\textbf{The $\Delta \beta$ photonic model.} ({\bf a}) Schematic diagram of the randomly varying propagation constants shown in different grayscales along the propagation direction of the seven-site structure. This set of random values is one example of the cases with a $\Delta \beta_A$ of 0.4~$\rm mm^{-1}$. ({\bf b}) The arrangement for the colored noise by the $\Delta \beta$ detuning for one site. ({\bf c}) The site energies for the seven sites of FMO and the base $\Delta \beta$ detuning used for the seven waveguides to match the site energies. 
}
\label{fig:strutturaChip}
\end{figure} 

\begin{figure*}[ht!]
\includegraphics[width=1.0\textwidth]{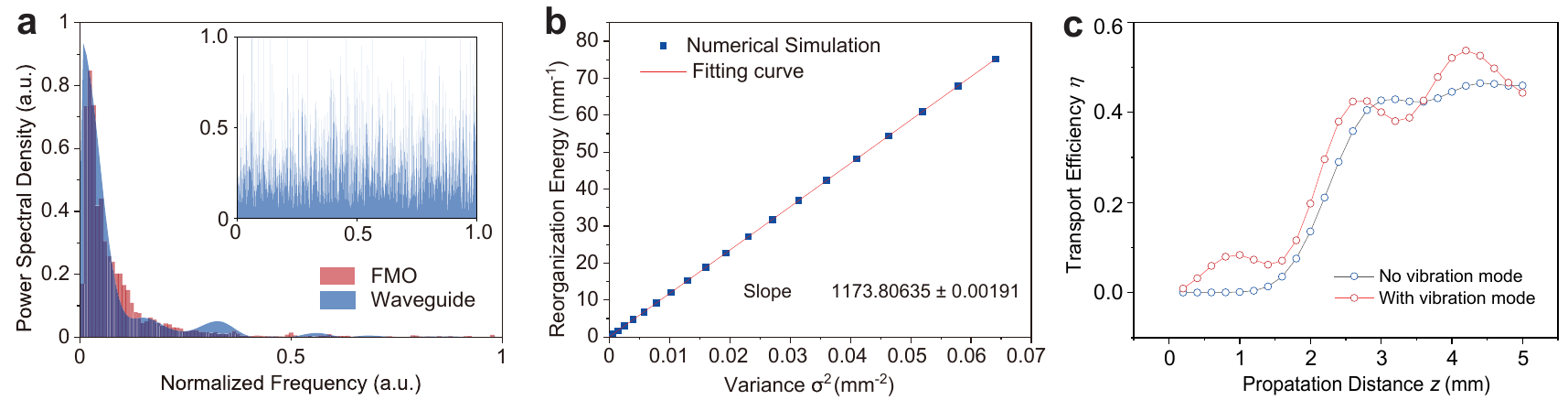}
\caption{\textbf{Photosynthetic energy transport using the $\Delta \beta$ photonic model.}({\bf a}) The spectral density formed by the colored noise introduced on Waveguide 3 is shown in the blue shadow. The red bars show the power spectral density for BChl 3 in a pigments of the monomeric subunits of the FMO protein via normal mode analysis from Ref\cite{Klinger2020}. Inset shows a spectral density generated from the white noise. ({\bf b}) Reorganization energy versus noise variance at site 7. The sampling frequency and period are $f_s = 1\ \rm mm^{-1}$ and $t_c = 20\ \rm mm$, respectively. The variance $\sigma^2(\Delta\beta_7) = \langle{(\Delta\beta_7)}^2\rangle - {\langle\Delta\beta_7\rangle}^2$ dependents on $\Delta\beta_7$ amplitude $\Delta\beta_{A,7}$. The parameters of the linear fitting are given in the bottom right of the figure. ({\bf c}) The transport efficiency at an early transport length with (red) and without (blue) vibrational assistance. $\Delta\beta_{A}$ for all waveguides are set to be 0.5 $\rm mm^{-1}$.}
\label{fig:apparato}
\end{figure*}

Using this photonic model, consider the case where each waveguide is broken up into many segments. We then have an effective piecewise dependent Hamiltonian:
\begin{equation}\label{Heff}
\begin{split}
H_{\rm {eff}}(t_{n})= \sum_{i}^N (\beta_{i0}+\Delta \beta_{i}(t_{n}))a_{i}^\dagger a_{i} \\+\sum_{j\neq i}^N C_{i,j} &(a_{i}^\dagger a_{j}+a_{j}^\dagger a_{i}),
\end{split}   
\end{equation}
where $\beta_{i0}$ are the base propagation constant for waveguide $i$, and $\Delta \beta_{i}(t_{n})$ is the extra detunings at segment $t_n$. Then we have the wavefunction:
$\Psi(t_n)=e^{-iH_{\rm {eff}}(t_{n})\Delta t}\Psi(t_{n-1})$,
where $\Delta t$ is the time interval for segment $t_n$.

Such an effective Hamiltonian $H_{\rm {eff}}$ can be straightforwardly mapped to the real modulation in segments of the photonic lattice. As shown in Fig.2a, we set the waveguide into segments of equal length with $\Delta t$ to be 1mm, and we introduce various random $\Delta \beta$ values ranging between 0 and a given amplitude denoted by $\Delta \beta_A$. At the end of the waveguide, measuring the light intensity distribution gives $|\Psi|^2$. 

In this experiment-friendly photonic model, we are able to introduce the noise consistent with the actual FMO complex. Instead of plain white noises, the biological energy transport involves colored noise that exhibits a non-Markovian nature \cite{Ishizaki2009}. Fig.2b shows an example on the $\Delta \beta$ values at Waveguide 7 that yield colored noises for Bchl7 (see details on generating white and colored noises in Supplementary Note 4 and 5). Besides, the base site energy for the seven Bchls in FMO varies. In order to reflect that, we additionally consider  \textcolor[rgb]{0,0,0}{different base values of $\beta_i$ for the seven waveguides representing the seven sites}, as shown in Fig.2c.

{\it The Power Spectral Density.} According to Wiener-Khinchin theorem, the power spectral density $J_{i}(\omega)$ is the Fourier transform of the original correlation function of the signal. In the context of the $\Delta \beta$ photonic model, it is   
\begin{equation}\label{PSD}
	J_{i}(\omega) 
= \int_{-\infty}^{+\infty} d\tau e^{-i\omega\tau}\langle\Delta\beta_i(\tau)\Delta\beta_i(0)\rangle.
\end{equation}
In Fig.3a, the red bars show the pattern of the intermolecular spectral density for BChl 7 of the actual FMO complex\cite{Wendling2000,Klinger2020}. \textcolor[rgb]{0,0,0}{The spectral density is obtained via normal mode analysis of
the whole pigment–protein complex with the charge density coupling method for the local optical transition energies of
the pigments by Klinger, $et~al$\cite{Klinger2020}.} It shows that the noise power for actual FMO complexes is concentrated in the low-frequency components, with some fluctuations in the mid-frequency region, and converges to 0 in the high-frequency region. Note that a hypothesis on site-independent spectral density is adopted for simplicity\cite{Cao2020}, and the distributions of power spectral density at the other six sites are in similar patterns\cite{Klinger2020}.

The blue shading area in Fig.3a represents the power spectral density for Waveguide 7 we generate using the colored noises via $\Delta \beta$ detunings. \textcolor[rgb]{0,0,0}{It matches the normalized pattern for actual FMO complex in terms of the shape on concentrating in low-frequency components.} On the other hand, in the inset Fig.3a we show a distinct pattern generated from the white noise. The white noise is actually Markovian while the colored noise exhibits strong non-Markovianity that does exist in the FMO complex\cite{Ishizaki2009, Chen2022}. See details on noise characteristics in Supplementary Note 6.

{\it The Reorganization Energy.} The photoexcitation is always accompanied by another energy transfer process. After the FMO complex is photoexcited to a localized excited state, the nuclei inside will undergo a relaxation process to achieve a new equilibrium position. The energy released during relaxation is characterized by the reorganization energy $E^R$\cite{Mancal2020,Cao2020}, which generally indicates the strength of system-bath coupling\cite{Mancal2020}. $E^R$ for each single site $i$ can be calculated by: 
\begin{equation}\label{reorganization}
	E_i^R = \frac 1 \pi \int_0^\infty d\omega \frac {J_i(\omega)} {\omega}.
\end{equation}

In the FMO complex, the reorganization energy follows a quantitative relationship with the variance of noise $\sigma^2$ \cite{Ishizaki2021}. For site $i$, there is 
\begin{equation}\label{linear}
	\sigma_i^2 = 2k_B T E_i^R,
\end{equation}
where $k_B$ is the Boltzmann constant. In our $\Delta \beta$ photonic model, $\sigma^2(\Delta\beta_i) = \overline{{\Delta\beta_i}^2} - {\overline{\Delta\beta_i}}^2$.
Varying the detuning amplitude $\Delta\beta_A$ from 0 to 1 $\rm mm^{-1}$, we get the corresponding $\sigma^2(\Delta\beta_i)$, and meanwhile, we work out $E_i^R$ by integrating the spectral density according to Eq.\ref{PSD} and \ref{reorganization}. As shown in Fig.3b, there is a high goodness of linear fit for the relationship between $E_i^R$ and $\sigma^2(\Delta\beta_i)$ in accordance with Eq.\ref{linear}.

{\it The Vibrational Assistance.} In up-to-date literature, the strong coupling to the the vibrational modes is believed to play an important role for energy transport\cite{Cao2020}. The electronic coherence, vibronic coherence and vibrational coherence are found to live with different coherence time, which are  50-100fs, \textless 500fs and \textgreater 1000fs, respectively\cite{Gelin2019, WangL2019}. Our 20-mm-long photonic chip corresponds to an evolution time in the magnitude of 10ps, which goes beyond the above time scale. Still, it is interesting to investigate a vibrational assistance at the very early stage of evolution. We set an additional vibrational mode by an extra waveguide, with a vibrational coherence strength equal to the gap between the two lowest eigenstates of the 7-site Hamiltonian. As shown in Fig. 3c, the transport efficiency is clearly enhanced when coupling to the vibrational mode than without such a coupling. Note that this example uses a $\Delta \beta_A$ of 0.5 $\rm mm^{-1}$ and for other noise amplitudes, the enhanced efficiency via vibrational assistance always exists. See more numerical details in the Supplementary Note 7.

Apart from the above issues, the issues on exciton transfer\cite{Rebentrost2009} and energy localization\cite{Coates2021} are also critical when discussing transport efficiencies. We illustrate the exciton transfer process inside FMO by numerically simulating the coherent light evolution in FMO-mimic waveguide array. By analyzing the most probable excited site when varying the propagation length, we see both large detunings of the site energy that induce strong disorders, and large-scale $\Delta \beta$ noises that induce the Zeno effect, would enhance exciton localization (See details in Supplementary Note 8). Furthermore, we show the disorder-induced enhancement of energy localization by simulating the energy transport efficiency and the distribution of eigen-energy levels (See details in Supplementary Note 9). 


\begin{figure*}[t!]
\includegraphics[width=0.85\textwidth]{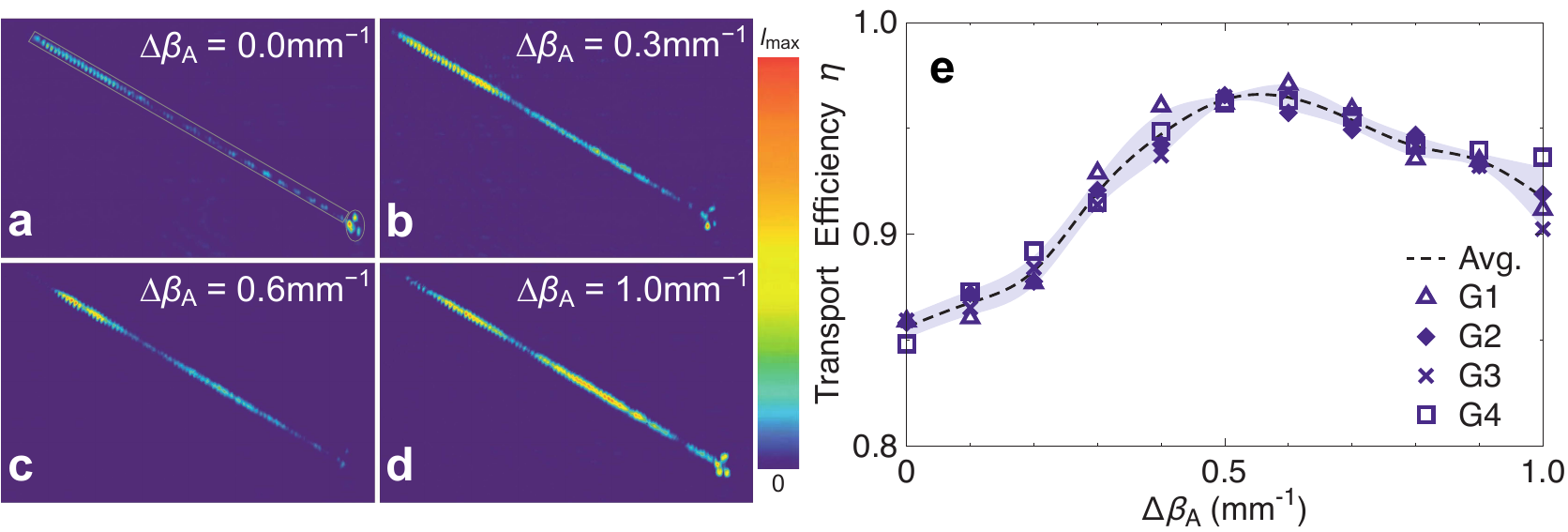}
\caption{\textbf{Experimental transport patterns and the energy transport efficiency.} Three examples from the arrays formed with different $\Delta \beta_A$ values show different energy transport efficiencies. The $\Delta \beta_A$ values are $0.0~\rm mm^{-1}$ for ({\bf a}), $0.3~\rm mm^{-1}$ for ({\bf b}), $0.6~\rm mm^{-1}$ for ({\bf c}) and $1.0~\rm mm^{-1}$ for ({\bf d}). The zones for seven-site FMO complex and the sink are marked with an ellipse and a rectangle respectively. The energy transport efficiencies are given in the bottom-left of the figures. ({\bf e}) The measured transport efficiencies for samples of different $\Delta \beta_A$ values. The results for each individual sample and the averaged values are plotted in dots and a curve, respectively.}
\label{fig:Results4}
\end{figure*}

{\it The Quantum Simulation Experiment.} In experiment, we have prepared four groups of samples in the \emph{colored} noise environment, having a range of $\Delta \beta_A$ values 0, 0.1, 0.2, ..., 1.0 $\rm mm^{-1}$. All samples have the noise settings that make a power spectral density consistent with the actual FMO complex. We inject the 810 nm vertically polarized coherent light into Waveguide 6 and measure the evolution patterns using a CCD camera. We are kind of simulating the energy packet random walk using a coherent light because when there is only one walker, the quantum coherence effects are well simulated by a coherent light field\cite{Jeong2004}. We process the figures to readout the light intensity in the seven-site part and the sink part, $I_{\text{FMO}}$, and $I_{\text{sink}}$, respectively. Then $I_{\text{sink}}/(I_{\text{FMO}}+I_{\text{sink}})$ is worked out as the energy transport efficiency $\eta$. In Fig.4a-d, the evolution patterns and the corresponding energy transport efficiencies for four samples in the same group of different $\Delta \beta_A$ values are presented. The experimental results for all samples are provided in Supplementary Note 10. 

We characterize the energy transport efficiencies for all samples (in Fig.4e) and show a prompt increase of the transport efficiency for all groups up to 96\% when $\Delta \beta_A$ increases up to 0.5-0.6~$\rm mm^{-1}$, followed up by an efficiency droop when $\Delta \beta_A$ further increases. Our experimental results based on the $\Delta \beta$ photonic model shows that the environment noise can assist quantum transport, just as the name ENAQT\cite{Rebentrost2009} suggests. We show an optimal $\Delta \beta_A$ that corresponds to an efficiency peak. This is consistent with our rich numerical analysis on the ENAQT effect (See Supplementary Note 11), where the optimal transport behavior always occurs at an intermediate dephasing scale.

In summary, we have fully explored the capabilities of the $\Delta \beta$ photonic model on quantum simulation of photosynthetic energy transport, with a case on the Fenna-Matthews-Olson complexes. By taking full advantages of the flexible arrangement of the array configuration in our three-dimensional photonic lattice, we manage a faithful layout of the coupling profile for the seven-site FMO complex. Meanwhile, the experimentally feasible $\Delta \beta$ tuning enables us to set the base site energy for different BChls in FMO, and build the non-Markovian colored noise that simulates the power spectral density of the actual FMO complex. Through these efforts, we experimentally demonstrate an optimal ENAQT transport in the photonic lattice. We also show that the $\Delta \beta$ photonic model not only simulates ENAQT theories extensively studied during 2010s, but also can address many up-to-date interesting topics related to FMO energy transport such as the vibrational assistance. 

\textcolor[rgb]{0,0,0}{Our quantum simulation experiment can be broadly adapted to simulating the photosynthesis processes in many other chlorophyll complexes, such as PE545, PE555, PC645, $etc$.\cite{Novoderezhkin2010, Chandrasekaran2016, Zech2014}, given their protein structures, Hamiltonian matrices and noise spectrum. Our quantum simulation experiments can give insights on the scale of noise modulation for energy transport in those chlorophyll complexes. This study may inspire applications for bioscience \cite{Choi2021}.  }

Our results demonstrate a powerful analog quantum simulator that can possibly be further applied to a rich diversity of researches on open quantum systems. We have noticed a recent work using digital quantum circuits to simulate the open quantum system dynamics \cite{Hu2022}. They simulate four sites for FMO complex and requires a long circuit depth, which will work on fault-tolerant devices in the future. The quantum simulation is indeed of a broad interest \cite{Georgescu2014} and strongly depends on the quantum hardware capabilities. Our integrated photonic chips have been used for various quantum simulation tasks\cite{Wang2019b, Wang2020, Tang2022b} in the noisy intermediate-scale quantum era, and many can be further turned into practical modules for quantum information processing modules, such as Haar random matrix generation\cite{Tang2022}, and on-chip quantum state preserving\cite{Tang2022b}, $etc.$ 
This work of simulating FMO complexes essentially constructs the non-Markovian environments in
photonic chips. Non-Markovian processes have been widely studied but still have many emerging new applications, $e.g.$, entanglement reactivation\cite{Pirandola2021}, quantum storage\cite{Andersson2019}. In order to carry out innovative exploration of non-Markovianity and its applications, strict experimental conditions are always required, and our experimental endeavor provides a new way out for implementing non-Markovianity.
The approach with flexible Hamiltonian mapping and controllable introduction of noise on integrated photonics is useful and worthy of further exploration on quantum simulation and quantum information processing applications.

\section{Data availability}
Data used for graphing in this paper are available from the corresponding author upon reasonable request.

\section{Acknowledgment} 
The authors thank Prof. Roberto Osellame and Prof. Jian-Wei Pan for helpful discussions. The authors thank Dr. Stephan Hoyer for discussing on the Hamiltonian matrix for the FMO complex, thank Dr. Kiran Khosla for discussing on the open quantum system theories, thank Dr. Song Ke for teaching the software on plotting the protein structure, and thank Prof. Fei Ma for sharing research works on vibrational assistance. This research was supported by the National Key R\&D Program of China (2019YFA0706302, 2019YFA0308700, 2017YFA0303700), the National Natural Science Foundation of China (61734005, 11761141014, 11690033, 11904229), the Science and Technology Commission of Shanghai Municipality (STCSM) (17JC1400403, 21ZR1432800, 22QA1404600), and the Shanghai Municipal Education Commission (SMEC) (2017-01-07-00-02- E00049). X.-M.J. acknowledges additional support from a Shanghai talent program. MSK's work is supported by the UK Hub in Quantum Computing and Simulation with funding from UKRI EPSRC grant EP/T001062/1 and a Samsung GRC grant.

\section{Author contributions}

H.T. conceived the experiments. X.-W.S, H.T., Z.-Y.S., Z.F., and X.T. fabricated the femtosecond laser direct-writing chips with colored noise sequence. X.-W.S. and T.-Y.W. performed the experiments. X.-W.S., H.-M.W., X.-Y.X. and J.G. collected and analyzed the data. H.T., X.-W.S. and R.S. conducted numerical simulations. X.-W.S., T.-S.H., and T.-Y.W. contributed to drawing the figures. H.T. and X.-W.S. developed the manuscript. M.S.K. and X.-M.J. contributed to the discussions of the results. X.-M.J. supervised the project. 

\section{Competing interests} 
The authors declare no conflicts of interest.

\clearpage



\pagebreak
\widetext
\begin{center}
\textbf{\Large Supplementary Materials}
\end{center}
\setcounter{equation}{0}
\setcounter{figure}{0}
\setcounter{table}{0}
\setcounter{page}{1}
\setlength{\parskip}{0.5\baselineskip}
\linespread{1.15}

\makeatletter
\renewcommand{\theequation}{S\arabic{equation}}
\renewcommand{\thefigure}{S\arabic{figure}}
\renewcommand{\bibnumfmt}[1]{[S#1]}
\renewcommand{\citenumfont}[1]{S#1}

\def\@hangfrom@section#1#2#3{\@hangfrom{#1#2#3}}

\titleformat{\section}{\raggedright\large\bfseries}{}{-1em}{}


\section*{Supplementary Note 1: The Hamiltonian matrix of the FMO molecule}

{The full Hamiltonian matrix of the FMO molecule of $C. tepidum$ is given in Eq.\ref{FMO}, unit in $\rm {cm^{-1}}$, according to previous calculations \cite{Adolphs2006, Hoyer2010}: 
{\begin{equation}\label{FMO}
H_{FMO} = 
\begin{pmatrix}
12410 &\textbf{-96.0} & 5.0 & -4.4 & 4.7 & -12.6 & -6.2 \\
 & 12530 & \textbf{33.1} & 6.8 & 4.5 & 7.4 & -0.3\\
 &  &  12210 & \textbf{-51.1} & 0.8 & -8.4& 7.6\\
 &  &  & 12320 & \textbf{-76.6} & -14.2 & \textbf{-67.0}\\
 &  &  &  & 12480 & \textbf{78.3} & -0.1\\
 &  &  &  &  & 12630 & \textbf{38.3}\\
 &  &  &   &  &  & 12440 \\
\end{pmatrix}.
\end{equation}}
}

{The off-diagonal part of the matrix shows the coupling profiles among the seven BChls. The sign of the coupling coefficients depends on the orientation of the transition dipoles since the coupling is treated as a dipole-dipole interaction\cite{Hoyer2016}, but the absolute values of the coupling coefficients are mostly concerned as the coupling strength between two sites. The off-diagonal numbers in bold show strong coupling coefficients beyond 15$\rm {cm^{-1}}$. They are strictly mapped on the photonic chip as shown in Fig.1 of the main text. The rest coupling coefficients are much lower, which have a minor influence on the coupling profile, and are not shown in the table in Fig.1. For these small FMO coupling coefficients, the coupling coefficients on chip may not be quantitatively proportional to the small ones in FMO, but qualitatively are consistent, that is, these coupling coefficients on chip are also small and do not influence the main coupling profile. }

{The diagonal part shows the different site energies for the seven BChls. The values are for $C. tepidum$ in trimer \cite{Adolphs2006, Hoyer2010}. BChl 3 clearly has the lowest site energy while BChl 6 has the highest. This is faithfully mapped on the seven waveguides using the $\Delta \beta$ photonic approach as shown in Fig.2 of the main text.}

\textcolor[rgb]{0,0,0}{Our quantum simulation experiment can be broadly adapted to the simulations of photosynthesis for many other chlorophyll complexes. Our quantum simulation experiments can give rich insights into the scale of noise modulation for the energy transport process, and compared with carrying out experiments on real chlorophyll organisms, it is much more feasible and accessible. As shown in Fig.\ref{fig:3complex}, given the protein structures for PE545, PE555, and PC645, as well as their Hamiltonian matrices \cite{Novoderezhkin2010, Chandrasekaran2016, Zech2014}, we show these structures can be mapped to the integrated waveguide array using our photonic model.}

\section*{Supplementary Note 2: Waveguide preparation}

All the waveguides are fabricated using the femtosecond laser direct writing technique\cite{Crespi2013, Chaboyer2015, Tang2018}. We direct a 513-nm femtosecond laser (upconverted from a pump laser of 10~W, 1026~nm, 290~fs pulse duration, 1~ MHz repetition rate) into a spatial light modulator (SLM) to shape the laser pulse in the temporal and spatial domain. We then focus the pulse onto a pure borosilicate substrate with a 50X objective lens (numerical aperture:~0.55). Power and SLM compensation were processed to ensure waveguide uniformity\cite{Tang2018}.

We have also characterized the quantitative control for $\Delta \beta$ detunings in the photonic lattice. We write one waveguide using a base speed $V_0$, and the other one using a different speed $V$ ($V-V_0=\Delta V$). In such a detuned directional coupler, the coupling mode method \cite{Szameit2007} gives the effective coupling coefficient, $C_{\rm {eff}}$, instead of $C_0$, the coupling coefficient for a normal directional coupler with $\Delta V=0$. $C_{\rm {eff}}$ contains the detuning effect from $\Delta \beta$ through\cite{Lebugle2015}: 
\begin{equation}\label{DeltaBeta}
C_{\rm {eff}}=\sqrt{(\Delta \beta/2)^2+C_0^2}.
\end{equation} 
Therefore, characterizing $C_{\rm {eff}}$ and $C_0$ gives $\Delta \beta$.


To characterize the $\Delta \beta$ quantitatively, we prepared 13 groups of samples with a base speed and a detuned speed corresponding to a $\Delta V$ from 0 to $30~\rm mm/s$. Each group has 12 samples of the same speed setting but different evolution lengths. Therefore, for each group, we characterize the intensity  intensity of the two modes, I(A) and I(B), of the directional coupler of one sample of a certain evolution length (Fig.S2a). We then fit the relationship between [$I(A)-I(B)$]/[$I(A)+I(B)$] and evolution length $z$ with a suitable sinusoidal curve to get $C_{\rm eff}$ or $C_0$ (when $\Delta V=0$), as shown in Fig.S2b. 

With these 156 samples, we then obtain $\Delta \beta$ (unit: $\rm mm^{-1}$) against $\Delta V$ (unit: $\rm mm/s$), as plotted in Fig.S2c. There can be a rough fit linearly: $\Delta \beta=0.02\times \Delta V$. Knowing this, we can randomly generate $\Delta \beta$ by setting a corresponding $\Delta V$. Our experimental limit for generating a $\Delta \beta$ can reach around $1.0~\rm mm^{-1}$ that requires a speed detuning of around $50~\rm mm/s$. This allows for rich manipulation of $\Delta \beta$ that is useful for quantum simulation on integrated photonic chips.

It is worth noting that an increase in writing speed in fact corresponds to a decrease in $\beta$. The sign matters for the manipulation of site energies, so for instance, for Site 3 which has a low site energy, we use a relatively high base writing speed. On the other hand, the dephasing terms by random $\Delta \beta$ detunings are not so much influenced by the singular $\Delta \beta$-$\Delta V$ relationship. The value of $\Delta \beta$ depends on how we regard the base value for $\beta$. For instance, a random series of $\Delta \beta$ follow a uniform distribution ranging between zero and a positive amplitude, $\Delta \beta_A$. We can also regard it ranges from negative $\Delta \beta_A$/2 and positive $\Delta \beta_A$/2, by leveraging the base by $\Delta \beta_A$/2 for all 7 sites. The absolute value for base $\beta$ does not matter. For instance, changing the base $\beta$ for all sites from 11660/mm to 0, the Hamiltonian evolution result will not change. On the other hand, the fluctuation scales caused by random $\Delta \beta$ detunings will do, and hence in Fig.4 of the main text, we compare the results using different strengths of $\Delta \beta_A$ from a weak value to a relatively stronger scale.

Therefore, in order to introduce the classical environment noise in the waveguide array of the seven-site system, we create the $\Delta \beta$ detunings on each of the seven waveguides, via detunings of the laser writing speed. For each waveguide, we vary $\Delta V$ 20 times in the 2.0-cm-long evolution length of each waveguide, that is, $\Delta V$ in each segment of 1 mm is constant, and the random $\Delta V$ values each waveguide follow a certain distribution under a given $\Delta V$ amplitude, that can yield the power spectral density in a similar shape to that for the actual FMO complex. 

{In addition, note that in this work, each waveguide representing the 7 FMO sites is made of short segments of different $\Delta\beta$ detunings, which is written by accurately setting the target laser writing velocity. How to accurately control the writing speed for writing a short segment length of 1-2mm? To make this work, we set the \emph{`setback-return'} strategy. Step 1: The laser beam accelerates to the target velocity; Step 2: The laser beam enters the chip and writes a segment of waveguide of the target velocity. Step 3: The laser beam immediately levers up after writing this short segment. Step 4: The laser beam moves on in the air to adjust to the target velocity for the next segment (let's call it Segment 2). Then the system sets back to the starting point of Segment 2 and the laser beam then lowers down to write the Segment 2. 

{With the `setback-return' strategy, we aim to make the laser beam only write in the chip when it's set as the target velocity for each short segment, while it does all the preparation work for achieving the velocity in the air. In practice, we need lots of characterization efforts to ensure it works. We need to set proper compensation for the set back distance in Step 4 in accordance with the conditions of the femtosecond laser writing machine. Otherwise, the segments might be separate from each other which fail to make a continuous waveguide, or the segments overlap with each other that deviates from the designed property of the waveguide. In previous works that have constant $\beta$ along the waveguide or changes $\beta$ at a longer segment length, the error in writing segments can be overlooked. In this work, however, this impact is vital as we have many short segments. We did characterization considering various conditions, and we observed the quality of each waveguide prepared in different conditions using a microscope. Eventually we gained the dataset of the parameters for the `setback-return' strategy that is suitable for this experiment.

\section*{Supplementary Note 3: Some discussions on $\Delta C$ }

The existence of $\Delta \beta$ also causes some fluctuations of the effective coupling coefficient denoted as $\Delta C(=C_{\rm {eff}}-C_0)$. Hence the off-diagonal modulation of the Hamiltonian is expected as well. A little algebra on the relationship between $C_{\rm {eff}}$ and $C_0$ according to Eq.\ref{DeltaBeta} gives: $(\Delta \beta/2)^2=C_{\rm {eff}}^2-C_0^2=(C_{\rm {eff}}-C_0)(C_{\rm {eff}}+C_0)$. Considering $\Delta C$ is relatively small, we approximate $C_{\rm {eff}}+C_0$ as $2C_0$, and we have $(\Delta \beta/2)^2 \approx \Delta C \times 2C_0$, which makes: $\Delta C \approx \Delta \beta^2/(8C_0)$.

We have considered replacing $C_{i,j}$ in Eq.(3) of the main text with $C_{i,j}+\Delta C$ to calculate the transport efficiency of the on-chip FMO structure against $\Delta \beta_A$ (Fig.S1a and S1b). As $\Delta C \approx \Delta \beta^2/(8C_0)$, the inclusion of $\Delta C$ causes a very marginal influence on the transport efficiency, since it scales quadratically with $\Delta \beta$ under a small scale of $\Delta \beta_A$. We can mainly consider the model with only diagonal $\Delta \beta$ terms, as given in Eq. (3). 

\section*{Supplementary Note 4: White noise generation}

{In the waveguide system, seven sets of white noise are independently produced for the detuning $\Delta\beta_i$ at seven FMO sites. Initially, we generate a set of random numbers in the interval $[0, \Delta\beta_A]$ as a time series $\Delta\beta(t)$ of white noise. The distribution of the white noise follows the probability density function (PDF)
\begin{equation}\label{whiterho}
	\rho(\Delta\beta)=
\begin{cases}
\frac{1}{\Delta\beta_A} ,& 0\leq\Delta\beta\leq{\Delta\beta_A}\\
0, &\rm otherwise
\end{cases}.
\end{equation}}

{The typical evolution time $t_c$ in the system is $0\sim100$ mm, and we modulate the detuning $20\sim 50$ times in a single waveguide. For example, if the length of the noise sequence is $n = 20$ and the total evolution time is $t_c = 40\ \rm mm$, then we consider the sampling frequency of the noise signal as}
{
\begin{equation}\label{fs}
	f_s = \frac n {t_c} = 0.5\ \rm mm^{-1}.
\end{equation}
}

{Now consider the correlation between two series of the noise signal. From a statistical perspective, the cross-correlation function (CCF) measures how closely two random signals match at different moments. Since the sampling frequency is sufficiently low, we employ the $sample$ CCF expression for discrete variables\cite{Box2008}}
{
\begin{equation}\label{rc}
	R_C(X,Y,\tau) = \frac {Cov(X,Y,\tau)} {Cov(X,Y,0)},
\end{equation}
where $Cov(X,Y,\tau)$ is the covariance of two series $X$ and $Y$
\begin{equation}\label{cov}
	Cov(X,Y,\tau) = E((X_{t + \tau}-\overline X)(Y_t - \overline Y)) = \frac {\sum_{t=1}^{n-\tau}{(X_{t+\tau} - \overline X)(Y_{t} - \overline Y)}} {n}.
\end{equation}
}

{On the other hand, the $sample$ auto-correlation function (ACF) can be represented as\cite{Box2008}
\begin{equation}\label{ra}
	R_A(X,\tau) = \frac {Acov(X,\tau)} {Acov(X,0)},
\end{equation}
}
{where $Acov(X,\tau)$ is the $estimated$ auto-covariance of a series $X$}
{
\begin{equation}\label{acov}
	Acov(X,\tau) = E((X_{t + \tau}-\overline X)(X - \overline X)) = \frac {\sum_{t=1}^{n-\tau}{(X_{t+\tau} - \overline X)(X_{t} - \overline X)}} {n}.
\end{equation}
}

{Note that the denominator in Eq.\ref{cov} and Eq.\ref{acov} is the length of the series $n$, not $n-\tau$, which may be valid in some other publications. }

{For the white noise, the covariance of detunings $\Delta\beta_i(t)$ and $\Delta\beta_j(t)$ is
\begin{equation}\label{whitecov1}
	Cov(\Delta\beta_i,\Delta\beta_j,\tau) = E((\Delta\beta_{i,t + \tau}-\overline{\Delta\beta_i})(\Delta\beta_{j,t} - \overline{\Delta\beta_j})) = E(\Delta\beta_{i,t + \tau}\Delta\beta_{j,t}) - E(\Delta\beta_i)E(\Delta\beta_j),
\end{equation}
where $\overline{\Delta\beta_i}$ is the mean value of time series $\Delta\beta_i$\cite{Box2008}. The expectation values $E(\Delta\beta_i)$ and $E(\Delta\beta_{i,t + \tau}\Delta\beta_{j,t})$ are calculated using Eq.\ref{whiterho} as
\begin{equation}\label{whiteexp1}
	E(\Delta\beta_i) = \int_\mathbb{R} \rho(\Delta\beta_i)\Delta\beta_i\,d\Delta\beta_i = \frac 1 2 \Delta\beta_{i,A}
\end{equation}
and
\begin{equation}\label{whiteexp2}
	E(\Delta\beta_{i,t + \tau}\Delta\beta_{j,t}) = \iint_{\mathbb{R}^2} \rho(\Delta\beta_{i,t+\tau},\Delta\beta_{j,t})\Delta\beta_{i,t + \tau}\Delta\beta_{j,t}\, d\Delta\beta_{i,t+\tau} d\Delta\beta_{j,t} = E(\Delta\beta_{i,t + \tau})E(\Delta\beta_{j,t}) = \frac 1 4 \Delta\beta_{i,A}\Delta\beta_{j,A},
\end{equation}
where the joint probability density function $\rho(\Delta\beta_i,\Delta\beta_j)$ is simply regarded as the product of two separate PDFs of independent random variances $\Delta\beta_{i,t + \tau}$ and $\Delta\beta_{j,t}$. Considering the amplitude $\Delta\beta_i$ is uniform over seven sites, the covariance of detunings $\Delta\beta_i(t)$ and $\Delta\beta_j(t)$ is 
\begin{equation}\label{whitecov2}
	Cov(\Delta\beta_i,\Delta\beta_j,\tau) = 0.
\end{equation}
}

{Adopting similar method, the ACF of a detuning $\Delta\beta_i(t)$ can be obtained. The auto-covariance }
{
\begin{equation}\label{whiteacov}
	Acov(\Delta\beta_i,\tau) = E(\Delta\beta_{i,t+\tau}\Delta\beta_{i,t}) - E(\Delta\beta_{i,t+\tau})E(\Delta\beta_{i,t}) = \frac 1 {12} {\Delta\beta_{A}}^2\delta_{\tau,0},
\end{equation}
where $\Delta\beta_A$ is the amplitude of energy fluctuation at all FMO sites. Note that if $\tau = 0$, there is
\begin{equation}
	E({\Delta\beta_{i,t}}^2) = \int_\mathbb{R} \rho(\Delta\beta_{i}){\Delta\beta_i}^2 \,d\Delta\beta_i = \frac 1 3 {\Delta\beta_{A}}^2.
\end{equation}
}
{
So the ACF of detuning $\Delta\beta_i$ is 
\begin{equation}\label{whitera}
	R_A(\Delta\beta_i,\tau) = \delta_{\tau,0}.
\end{equation}
}
{In the following, we only consider the noise characteristics within one single waveguide detuning for there are no interesting relations between two different waveguides. }


\section*{Supplementary Note 5: Colored noise generation}


{Generally, the energy fluctuation in FMO environment is not uniform in the frequency domain, but has a featured pattern. To effectively simulate the energy transport in this environment, we additionally produce a type of colored noise by setting a response function. }

{Firstly, we generate a series of values from a normal distribution with mean 0 and standard deviation 1, and regard the pseudorandom numbers as a Gaussian white noise sequence picked with frequency $f_s$ during the time period of $t_c$. Then input the Gaussian white noise into a system described by the response function
\begin{equation}\label{filter}
	F(s) = \frac 1 {10s + 1} + \frac 1 {100s^2 + 10s + 1}.
\end{equation}
The colored noise is acquired by taking the absolute values of the output and normalizing the amplitude. }

{Actually, the colored noise we employed does not respect a specific distribution function, like the standard Gaussian or Lorentzian distributions. In the analog method, the time response of a dynamic system to random number inputs guarantees the dependence on historical states and preserves a considerable amount of randomness. In light of this, we directly perform numerical simulation of colored-noise ACF in the next section, instead of analytical derivations.}

\section*{Supplementary Note 6: Noise characteristics}

{According to Wiener-Khinchin theorem, the PSD is the Fourier transform of the original correlation function of the signal, which is
\begin{equation}\label{PSD}
	J_{i}(\omega) 
= \int_{-\infty}^{+\infty} d\tau e^{-i\omega\tau}\langle\Delta\beta_i(\tau)\Delta\beta_i(0)\rangle.
\end{equation}
Substituting Eq.\ref{whitera} into Eq.\ref{PSD}, we obtain the PSD of the white noise at a single site as
\begin{equation}\label{whitePSD}
	J_i(\omega)=\int_{-\infty}^{+\infty} d\tau e^{-i\omega\tau}\frac 1 {12} {\Delta\beta_{A}}^2\delta(\tau)=\frac 1 {12} {\Delta\beta_{A}}^2,
\end{equation}
where $\delta(\tau)$ is the Dirac delta. Eq.\ref{whitePSD} exhibits that the power spectrum of white noise is uniform for an untruncated time series. }

{To estimate the power spectral density (PSD) of the finite noise sequence utilized in simulations and experiments, we adopt the one-side periodogram method with a fast Fourier transform (FFT) of length 128. In our waveguide structure, seven noise sequences $\Delta \beta_i\ (i = 1,2,...,7)$ are obtained independently for seven sites. That is, we assume that the noise at each site is uncorrelated with each other. }

{The estimated PSD of white noise has been illustrated in the inset of Fig.3a in the main text.  
In this simulation, the total length of the noise series is 100, which is already enough to observe the uniform distribution of noise power in the frequency domain. In this perspective, white noise cannot effectively model the influence of environmental fluctuation on the energy transport of FMO complex. }

{In one simulation with sampling frequency $f_s = 0.5\ \rm mm^{-1}$ and total evolution time $t_c = 40\ \rm mm$, the estimated PSD of the noise at one site has been shown in Fig.3a, 
corresponding to the $\Delta \beta$ configuration shown in Fig.2b, both of the main text. It can be seen from Fig.2b 
that the noise at time $t + 1$ is in the vicinity of the noise value at time $t$, which exhibits the feature of history dependence. }

{Furthermore, the pattern of PSD is a simulation of the intermolecular spectral density of actual FMO complex\cite{Markus2000,Alexander2020}. Note that a hypothesis that the spectral density is site-independent is adopted for simplicity\cite{Cao2020}, and the distributions of power density at the other six sites are in similar patterns. It is clear that the noise power is mainly concentrated in the low-frequency components, and the maximum power appears around 0. But there are still some fluctuations in the mid-frequency region, leading to several extreme values of power. In the high-frequency region, the noise power converges to 0.}



{For the interest of specifying the non-Markovianity of the system, the sample ACFs of both white noise and colored noise are evaluated using Eq.\ref{ra} and Eq.\ref{acov}. In Fig.\ref{fig:Results8}, the horizontal coordinate $lag$ indicates the delay time $\tau$ set in Eq.\ref{ra}. If $\tau = k$, the vertical coordinate indicates the so-called $k$th lag ACF. In our calculations, the lag (mm) must be less than $t_c - 1/f_s$. Note that the evolution time $t$ is replaced with propagation length $z$ and the $k$th lag is for the detuning $\Delta\beta$ at distance $k / {f_s}$. }


{Fig.\ref{fig:Results8}b shows the sample ACF of white noise with sampling frequency $f_s = 0.5\ \rm mm^{-1}$ and sampling period $t_c = 100\ \rm mm$. The white-noise ACF rapidly decays as lag increases, since only the auto-correlation at lag $\Delta z = 0$ has a value around 1 $\rm mm^{-2}$. This phenomenon implies a weak pre-post correlation in the time series, and the system is of great Markovianity. }

{Fig.\ref{fig:Results8}a shows the sample ACF of colored noise with sampling frequency $f_s = 0.5\ \rm mm^{-1}$ and sampling period $t_c = 100\ \rm mm$. The largest auto-correlation $1\ \rm mm^{-2}$ emerges at $0$th lag, and the ACF oscillates slowly as lag increases. Notice that even $10$th lag ACF has the value around $0.4\ \rm mm^{-2}$. Contrary to white-noise ACF, a longer anto-correlation time scale appears in the colored-noise system. In this perspective, there exists a non-Markovian nature in the colored-noise structure. We suppose the fluctuation of the environment is not so fast (corresponding to a smaller decay rate), and the state of the noise environment at time $z$ has a certain degree of impact on the behavior of the system in the next moments over the period of auto-correlation time. }

\section*{Supplementary Note 7: Vibrational assistance}

{The intermolecular vibrational assistance is modeled by adding an extra waveguide, which is regarded as site 8. For the Hamiltonian matrix with $C_{chip}$ and site energies $\beta_i$ as shown in Fig.1c and Fig.2c of the main text, we can calculate the eigenvalues of the 7$\times$7 Hamiltonian matrix. The energy gap between the ground state and the first excited state is 0.4776 $\rm mm^{-1}$. To achieve the oscillation between these two sites and Site 8, the coupling coefficients between the 8th waveguide and the other seven waveguides are all set to 0.4776 $\rm mm^{-1}$.}


{Our photonic chip has an evolution length of 20mm. Considering a refractive index of around 1.5 and a proportionally elongated time as $C_{chip}$ is 14\% of $C_{FMO}$, this chip length corresponds to an evolution time for actual FMO energy transport in the magnitude of 10ps. Since the vibrational coherence occurs in a short time scale, we numerically simulate vibrational assistance at the onset of the evolution. In Fig.3c of the main text, we have shown the simulated energy transport efficiency for samples of a $\Delta \beta$ amplitude of 0.5 $\rm mm^{-1}$. It shows the energy transport efficiency with the vibration mode undergoing an oscillatory climb, while in the curve of energy transport without vibrational assistance, this pattern is not evident. In Fig.\ref{VibAssist}, we provide a more general analysis on vibrational assistance in samples of different noise strengths, $i.e.$, different $\Delta \beta$ amplitudes ranging from 0 to 1.0 $\rm mm^{-1}$ and all in the colored noise configuration consistent with previous sections. We show that the oscillatory behavior and enhanced energy transport efficiency at the very beginning always occur when the vibration mode is included.}

\section*{Supplementary Note 8: Discussion on exciton transfer }

\subsection*{A. Derivation on energy transport efficiency and transfer time via exciton recombination and trapping}

In photosynthetic complexes where energy is transferred by excitons, the energy transport efficiency and transfer time can be expressed in terms of the exciton recombination rate $\Theta$ and exciton trapping rate $\kappa_m$ at each site\cite{Mohseni2008, Rebentrost2009}. The two processes are represented by the anti-Hermitian Hamiltonian part\cite{Mohseni2008}
\begin{equation}
\begin{aligned}
H_r &= -i\hbar \Theta \sum_m |m\rangle\langle m|, \rm and\\
H_t &= -i\hbar \sum_m \kappa_m | m\rangle\langle m|.
\end{aligned}
\end{equation}
The corresponding energy transfer efficiency and transmission time are respectively denoted as\cite{Rebentrost2009}
\begin{equation}\label{2009Rebentrost8&9}
    \begin{aligned}
        \eta &= 2\sum_m\kappa_m\int_0^\infty {\rm d}t\langle m|\rho(t)|m\rangle\\
        \tau &= \frac 2 \eta\sum_m \kappa_m\int_0^\infty {\rm d}t\cdot t\langle m|\rho(t)|m\rangle,
    \end{aligned}
\end{equation}
where the transfer time can be recognized as a weighted average over the time domain.

In the waveguide system, the wave function at time $t$ is
\begin{equation}
    \begin{aligned}
    |\psi(t)\rangle &= \mathcal{T} e^{-i\hbar\int_0^t H(t'){\rm d}t'}|\psi(0)\rangle\\
    &= e^{-iH_n\Delta t}e^{-iH_{n-1}\Delta t} ... e^{-iH_2\Delta t}e^{-iH_1\Delta t}\cdot |\psi(0)\rangle,
    \end{aligned}
\end{equation}
where $\mathcal T$ is the Dyson chronological operator, $|\psi(0)\rangle$ is the initial state and $\{H_1, H_2, ..., H_n\}$ is a sequence of Hamiltonian containing different dephasings in the time range $(0, t)$. Each waveguide is an eigenmode with an energy transfer efficiency from waveguide 3 to waveguide $m$ as ${|\langle m | \Psi(t)\rangle|}^2$. Let the energy transport efficiency at time $t=j\delta t$ be $\eta_j$, then the percentage of energy transferred in time $(t, t+\delta t)$ is $\eta_{j+1}-\eta_j$. Corresponding to Eq.\ref{2009Rebentrost8&9}, the transfer time can be reformulated in the waveguide system as 
\begin{equation}
    \begin{aligned}
    \tau &= \frac 1 {\eta_N} \sum_{j=1}^N \delta \eta_j \cdot j\delta t\\
    &= \frac 1 {\eta_N} \sum_{j=0}^N (\eta_{j+1}-\eta_j) \cdot j\delta t\\
    &= \frac 1 {\eta_N} \sum_{j=0}^N \left[\sum_{m=8}^{87}{|\langle m | \Psi((j+1)\delta t)\rangle|}^2-\sum_{m=8}^{87}{|\langle m | \Psi(j\delta t)\rangle|}^2 \right] \cdot (j+\frac 1 2)\delta t\\
    &= -\frac {\delta t} {\eta_N} \sum_{j=1}^{N-1} \sum_{m=8}^{87}{|\langle m | \Psi(j\delta t)\rangle|}^2 + (T - \frac 1 2 \delta t),
    \end{aligned}
\end{equation}
where the total transfer time is $T = N\delta t$, and $\delta t\ll \Delta t$ is satisfied.


In some cases, integrating the above original equation from zeros to infinity yields dip\cite{Rebentrost2009}. However, for the 20-mm-long photonic waveguide, although the transport efficiency has a peak, the limited truncation time and the bidirectional coupling between Site 3 and the sink lead to a transfer time where no obvious dip is observed. That is, within the experimental time range, the structures with low transport efficiency will maintain a certain light intensity for some time, and the structures with high transport efficiency perform better all the time, so the latter always have more weight in comparison. Therefore, while in biological systems, the definition of transport efficiency and transfer time does closely relate to exciton capture and exciton complexation rate, in this experiment, the definition of average transfer time does not provide as much information as transport efficiency. We hence mainly investigate exciton transfer via transport efficiency.


The above discussion is about how FMO's biological concepts are combined with waveguide systems. The mapping between observable measurements of waveguide systems and FMO excitons energy transfer is given below.

\subsection*{B. Relation between waveguide light transfer and FMO energy transfer }

\emph{Single photon.}
If a single photon evolves in a waveguide array, there is only one photon being measured in one waveguide at the outgoing end. This is a probabilistic collapse process. The probability pattern in collapse is determined by the waveguide structure, $i.e.$, the waveguide Hamiltonian matrix. By increasing the number of repeats of the single photon injection and the measurement, the observed value of probability can gradually approach the real value.

\emph{Coherent light.}
It is theoretically proved that the results of single-photon quantum walks have the same probability distribution as those of continuous laser\cite{Jeong2004}.
In our experiments, the distribution of the light intensity at the outgoing end directly represents the probability distribution. The simulation of the FMO energy can be calculated according to 
\begin{equation}
    \begin{aligned}
        E_{7-site} &=  \sum_{i=1}^7 p_i\epsilon_i,\\
        \eta &= \frac {E_{sink}} {E_{total}} = \frac {\sum_{i=8}^{87}p_i\epsilon_i} {\sum_{i=1}^{87}p_i\epsilon_i} = \frac {\epsilon_{sink}\sum_{i=8}^{87}p_i} {E_0}\\
        &= \frac {\epsilon_{sink}\sum_{i=8}^{87}p_i} {\epsilon_6 \times 1} = \frac {\epsilon_{sink}} {\epsilon_6} \times \sum_{i=8}^{87}p_i\\
        &= \frac {\epsilon_{3}} {\epsilon_6} \times \frac {I_{sink}} {I_{total}}.
    \end{aligned}
\end{equation}
Here, $p_i$ is the probability of measuring the photon at site $i$, and $\epsilon_i$ is the site energy of site $i$. We use the energy conservation condition to obtain $E_{total}=E_0$, where $E_0$ is the total energy at the initial time. For sink waveguides, we all adopt the same laser-writing speed as waveguide 3 to ensure the site energy $\epsilon_{sink}=\epsilon_3$. $I_{sink}$ and $I_{total}$ are respectively the light intensity detected at the sink part and the whole waveguide array. 

The above discussion demonstrates the rationality of light intensity measurements. The following context provides information on single-site excitation probability corresponding to the single-exciton behavior.

\subsection*{C. Excitation site evolution}

From the above discussion, it can be concluded that the single-site excitation inside the 7-site waveguide structure is not related to the measurement of FMO energy transport efficiency, although the exciton transfer inside the 7-site structure does have a significant impact on the way the energy flows from Site 6 to Site 3 and then the sink part.

For the purpose of illustrating the energy flow pattern in 7-site structure, we numerically calculate the light intensity of a single site at different moments, $i.e.$, the probabilities of measuring an excited site varying with time evolution. This is the result of a single-excitation manifold.

We first consider the scenarios for varying the disorder strength, which refers to the static change of the site energy represented by $\beta$. 
The evolution of normalized probability of a 7-site structure with varied disorder strengths is shown in Fig.\ref{fig:Results17}, where different colors indicate different excitation sites. From top to bottom, different color blocks indicate the probability of excitation from Site 7 to Site 1. The sum of the normalized probabilities of different sites satisfies the normalization condition $\sum_{i=1}^7p_i=1$. At the initial moment, only Site 6 has an excitation probability of 1, representing laser injection only from waveguide 6. The triangles mark one of the seven sites most likely to be excited. From Fig.\ref{fig:Results17}a to d, the order of disorder strength $\Gamma$ is 0, 3, 6 and 10 $\rm mm^{-1}$. All subgraphs have a noise intensity of 0.

The change of the most probable excited site illustrates the path of exciton energy transport to some extent. The outlet of energy transfer is at site 3, which is directly connected to the sink. Site 4 has a very close connection to site 6. With the increase of disorder intensity, exciton is more likely to be detected at Site 4 or Site 6.

Besides, we work on the scenarios with varying pure dephasing strength, represented by the amplitude for the random detunings, $\Delta\beta_A$.
The evolution of normalized probability of 7-site structure with varied pure dephasing strength is shown in Fig.\ref{fig:Results18}. From Fig.\ref{fig:Results18}a to f, the noise intensity $\Delta\beta_A$ is 0.1, 0.3, 0.5, 0.7, 1.0 and 80 $\rm mm^{-1}$. All subgraphs have a disorder intensity of 0. 
In these simulation results, Fig.\ref{fig:Results18}c and Fig.\ref{fig:Results18}d with noise amplitude 0.5 $\rm mm^{-1}$ versus 0.7 $\rm mm^{-1}$ have the most dispersed exciton distribution. In the main text, we treat the system slightly differently, by setting the FMO-mimic waveguide array as a whole and considering a small disorder strength that maps FMO. Still, we experimentally measure an optimal noise amplitude of about 0.6 $\rm mm^{-1}$ corresponding to the highest transmission efficiency. Here this simulation in Fig.\ref{fig:Results18} is consistent with our experimental result. 

In Fig.\ref{fig:Results18}f,  the noise is getting too strong, it is equivalent to applying a high-frequency measurement. The system exhibits the Zeno effect, that is, the original 2mm evolution process (Fig.\ref{fig:Results18}a) needs 20mm (Fig.\ref{fig:Results18}f) to complete. 
Note that the localization due to disorder has different dynamics with the Zeno effect due to strong noise (dephasing), although both result in a low energy transport efficiency.

\section*{Supplementary Note 9: ENAQT in waveguide system}
\textcolor[rgb]{0,0,0}{Fluorescence resonance energy transfer (FRET) was proposed in 1948 by F{\"o}rster, also known as F{\"o}rster resonance energy transfer \cite{Forster1948}. FRET was first experimentally validated in 1967 \cite{Stryer1967}. To date, many new techniques have been derived from FRET. FRET describes the excitation of a donor molecule that jumps from a ground state to an excited state while generating an oscillating dipole that resonates with the dipole of the acceptor molecule. The energy transfer from the donor to the acceptor is achieved by dipole-dipole interactions and hence FRET is a non-radiative energy transfer process. FRET is characterised by the fact that the transfer efficiency is inversely proportional to the sixth power of the distance between the donor and the acceptor. The energy transfer efficiency of FRET depends mainly on the overlap of the donor's emission spectra with the acceptor's absorption spectra, the spacing between the donor and the acceptor, and the orientation of the dipole moments. The energy donor and acceptor of FRET can be organic small molecules or biological macromolecules, offering the possibility of modelling energy transfer in living organisms. }

\textcolor[rgb]{0,0,0}{Although FRET achieved some interpretability early on, as research progressed, it was found that coherence and decoherence play an important role in the energy transfer of the complex. In 2016, Park et al. investigated the energy transfer efficiency in the M13 virus and demonstrated that decoherent quantum walks predicted the energy transfer better than F{\"o}rster theory \cite{Park2016}. This suggests that the classical random walk dominated by the classical master equation is flawed in its description of diffusion, and in particular cannot predict strong coupling with large chromophoric arrays. Therefore, it is the decoherent quantum walk dominated by the quantum master equation that is an effective tool for predicting exciton energy transfer in complexes.}

\textcolor[rgb]{0,0,0}{In contrast to classical or semi-classical theories such as FRET, ENQAT theory takes into account decoherence due to noise and the noise spectral density that can be controlled for non-Markovianity. This provides us with favourable tools to simulate energy transport suitable for specific complexes by shaping the spectral density.}

Environment-assisted quantum transport is well known for playing an important role in living organisms, especially in photosynthetic energy transfer process, although recent studies suggest it may work in more subtle ways\cite{Plenio2008,Mohseni2008,Lambert2013,Kassal2013,Engel2007,Chin2010,Harush2021,Stones2016,Higgins2021,Duan2017}. In this section, we illustrate ENAQT in waveguide systems, showing that the FMO-mimic structure is affected by both static disorder and dephasing, and that the experiments are a search for optimal dephasing in the weak disorder case.

\subsection*{A. Simulation on ENAQT}


To illustrate that the peak of optical intensity transmission efficiency occurs in this system as a result of ENAQT, we evaluate the degree of localizability and observe the effect of dephasing on transport efficiency at different disorder intensities. Among the 87 waveguides in this system, only the first 7 waveguides are used to build the FMO structure. Although it is theoretically possible for ENAQT to occur in larger subsystems, only the first 7-waveguide characteristics have a determinate impact on the simulation of the FMO energy transport efficiency. Recall that the waveguide Hamiltonian without dynamic $\Delta\beta(t)$
\begin{equation}\label{Maintext1}
H_{\rm {eff}}= \sum_{i}^N \beta_{i0}a_{i}^\dagger a_{i} +\sum_{j\neq i}^N C_{i,j} (a_{i}^\dagger a_{j}+a_{j}^\dagger a_{i}), 
\end{equation}
where $\beta_{i0}$ is the constant propagation coefficient of a single waveguide. By adding the static disorder $\gamma_i0$ to Eq.\ref{Maintext1}, the Hamiltonian of the waveguide can be expressed as
\begin{equation}
H'_{\rm {eff}} = \sum_{i}^N (\beta_{i0} + \gamma_{i0})a_{i}^\dagger a_{i} +\sum_{j\neq i}^N C_{i,j} (a_{i}^\dagger a_{j}+a_{j}^\dagger a_{i}).
\end{equation}
For simplicity, the disorder strength of each waveguide is set equal and the individual disorders are sampled in a uniform distribution, $i.e.$, $\gamma_{i0}\sim U(0, \Gamma)$. Many studies have shown that introducing static disorder to a many-body system can lead to an enhancement of localization\cite{Anderson1958,Coates2021}. Here, we use the inverse participation ratio (IPR) to quantify the localizability. The expression of IPR is\cite{Coates2021}
\begin{equation}
	{\rm{IPR}} = \frac 1 {\sum_{i,\alpha}{|\langle i |E_\alpha\rangle |}^4},
\end{equation}
where $|i\rangle$ and $|E_\alpha\rangle$ are respectively the single-excited state at site $i$ and the $\alpha$-th eigenstate, and the denominator is summed over all $i$s. Due to the coupling between neighbor waveguides, an eigenstate is generally a superposition of several waveguide-excited states, i.e., a single waveguide does not correspond to an eigenstate. Considering the normalization condition $\sum_i{|\langle i | E_\alpha\rangle|}^2 = 1$, it is clear that smaller IPR implies the enhanced localization of the 7-site subsystem.

The relationship between localizability and disorder is shown in Fig.\ref{fig:Results15}a. The solid blue line indicates the transport efficiency and the red dotted line indicates the IPR. Shading is the standard deviation obtained after 1000 samples. As seen from the figure, both transport efficiency and IPR experience a decrease when the static disorder intensity increases.

To further show the effect of the dynamic $\Delta\beta(t)$ on the energy transfer of the system with different localizability, we add the time-dependent $\Delta\beta(t)$ at the asterisk, and the results of the numerical simulation are sketched in Fig.\ref{fig:Results15}b-d, where the shading is the standard deviation obtained after 100 samples. At different disorder amplitudes, there are always peaks caused by dephasing. The $\Gamma=0$ corresponds to our experimental condition, i.e., only the natural site energy of FMO complex. Here, we explore a wider parameter space. On both sides of the peak, the decrease in energy transfer efficiency is not identical. In the absence of dephasing, the energy still maintains a certain transfer efficiency, which benefits from the fact that the dynamics of the system are mainly affected by static disorder at this time, but the disorder $\beta_{i0}$ caused by the FMO site energy alone is weak. Therefore, the left side demonstrates that the weaker disorder (in the case of $\Gamma = 0$) causes a certain localization. However, when the dephasing is too strong, the super-strong time-dependent $\Delta\beta$ is equivalent to performing high-frequency probing, confining the system dynamics in Zeno subspace. Fig.\ref{fig:Results15}b-c show the ENAQT at $\Gamma=10 \rm mm^{-1}$ versus $\Gamma=100 \rm mm^{-1}$, with peak at $\Delta\beta_A=2.4\rm mm^{-1}$ versus $\Delta\beta_A=23.6\rm mm^{-1}$, respectively. This phenomenon is consistent with the physical pattern that disorder enhances localizability and thus requires more dephasing to break the energy transfer barrier\cite{Caruso2009,Chin2010,Zerah2020,Coates2021}.

It should be noted that, according to the measured curves of the experimental parameters, many regions in Fig.\ref{fig:Results15} are beyond existing experimental conditions. This is because the required femtosecond laser direct writing speed is too high. In the range of smaller parameters, we display the agreement of the numerical simulation with the experimental measurements (see main text).

\subsection*{B. Simulation on localizability}

The sources of ENAQT can be roughly divided into the expansion of energy levels by dephasing, noise-assisted energy outflow of invariant subspace, and noise-assisted group velocity redistribution in momentum rejuvenation.

In Fig.\ref{fig:Results15}a, IPR under different disorder intensities is calculated. Here we attempt to further uncover the source of ENAQT in this structure. Fig.\ref{fig:Results16} shows the eigenenergy distribution under different disorder intensities. From Fig.\ref{fig:Results16}a to f, the disorder intensities $\Gamma$ is 0, 3, 6, 10, 20 and 100 $\rm mm^{-1}$ in order. In Fig.\ref{fig:Results16}, the size of the bubble represents the probability of being in that eigenstate. In the 7-site structure, there are seven different eigenenergy levels corresponding to seven different eigenstates, which means no degenerate subspace is observed. At each site, the probability sum of being in all energy levels is 1, which is the requirement of normalization.

The following information can be found in Fig.\ref{fig:Results16}:

1) With the increase of $\Gamma$, the range of energy distribution becomes wider. For example, the energy of the highest excited state in a is 4 $\rm mm^{-1}$ higher than that of the ground state in Fig.\ref{fig:Results16}a, but this gap is 83 $\rm mm^{-1}$ in Fig.\ref{fig:Results16}e, which means that the band gap between other energy levels is increased.

2) As $\Gamma$ increases, some energy levels move closer to each other, while others move away. The energy level distribution in Fig.\ref{fig:Results16}a is more uniform, while the energy gap between the 4th and 5th excited states in Fig.\ref{fig:Results16}f is 2 $\rm mm^{-1}$, which is significantly different from the 22 $\rm mm^{-1}$ energy gap between the 5th excited state and the highest excited state.

3) With the increase of $\Gamma$, one site is more and more concentrated in a single energy level. Up to the case in Fig.\ref{fig:Results16}f, the probability of measuring a particular eigenenergy at one site is about 1.

The above characteristics show that disorder does change the distribution of energy levels and eigenstates of the system. In systems with more disorder, the energy gaps are more inconsistent, resulting in more localization.

\section*{Supplementary Note 10: Experimental results on energy transport}

{In Fig. \ref{Group1}-\ref{Group4}, we show all the experimental evolution patterns and the measured transport efficiencies for the four groups of samples. The samples have colored noises, and the base $\Delta\beta$ follows Fig.2c to map the site energy for actual FMO complexes. For each group of samples, the noise configurations are similar, except that the absolute strength scales according to the given $\Delta \beta_A$ ranging from 0.1 to 1.0 $\rm mm^{-1}$. } 

\textcolor[rgb]{0,0,0}{In Fig. \ref{FigureS21}, we display the experimental measured energy transfer efficiency of FMO-mimic waveguide system with white noise. All measured points are illustrated in Fig. \ref{FigureS20} with the fitted curve. One can see that the optimum noise amplitude is around 0.3 $\rm mm^{-1}$, which differs from the FMO-mimic waveguide with colored noises shown in Fig.4 of the main text. To shed light on the effect of this noise nature on the FMO energy transfer, we give further numerical analyses in Supplementary Note 11.}


{When collecting the data from experiments, we obtained the corresponding ASCII file, which is essentially a matrix of pixels. We created a `mask’ that highlights the pixels in an ellipse that contains all the light intensity from the waveguides representing seven-site FMO, and another mask in a rectangular shape for all the light intensity from the waveguides representing the long sink. We then summed the light intensity for all the pixels within each mask using Matlab. The normalized proportion of light intensity for each circle represents the probability at the corresponding waveguide.}

\section*{Supplementary Note 11: Simulation on energy transport efficiencies and the influence on noise distribution}

{We simulate the energy transport processes influenced by the colored noise with different amplitudes and different settings of the segment length. Fig.\ref{fig:Results6}a illustrates an energy transport process versus the propagation distance. It shows that the transport rate decreases with increasing propagation distance and no oscillations emerge. At a propagation distance of 20mm, it shows a relatively stable transport efficiency. Hence it is suitable that in experiments, we take a view at a propagation length of 20mm to compare their transport efficiencies.}

{Fig.\ref{fig:Results6}b shows the transport efficiency using different segment lengths, namely, turning 20mm evenly into 10, 20, 40, 60 or 80 segments, respectively. It turns out that too frequent change of $\Delta \beta$ does not necessarily make a fast energy transport, and the setting of 20 segments is a favorable choice with high transport efficiency. This is consistent with our previous work on quantum stochastic walks on photonic chips\cite{Tang2022}. In Fig.S13 of that work, we show that there exists a preferable segment length that leads to faster convergence to the Haar measure. Here we use a segment length of 1mm, not too long or too short, considering both the need for enough segment length to reach a high enough transport efficiency, and the need for enough segment counts to form a clear power spectral density. }

{In this section, we further discuss how noise of different functional characteristics can affect energy transport efficiency. The numerical simulation results show that the distribution profile of the noise does not have a decisive influence on the energy transport efficiency. In contrast, the intensity of the noise has a major influence on the optimum energy transport efficiency, which can be revealed by the power spectral density. This pattern ensures the validity of the results of our FMO waveguide experiments, even if our ambient spectral density is slightly distorted from the real chemical environment of the FMO, which is limited by the precision of the femtosecond laser direct writing optical waveguide. }

{Here we define five types of noise distribution:
}

\subsection*{A. Uniform distribution}

{
If the random variable $X$ has a probability density function
\begin{equation}\label{uniform1}
	p(x)=
\begin{cases}
\frac{1}{b-a}, & a\leq x\leq b,\\
0, & \rm otherwise.
\end{cases}
\end{equation}
then $X$ is said to obey a uniform distribution on the interval [a,b], denoted $X\sim U[a,b]$. The fluctuation of site energy $\Delta\beta_i\sim U[0,1]$.
}

{
We scale the noise amplitude to demonstrate optimal energy transport efficiency. However, the number of segments is limited to a maximum of 20 and the random number generation is subject to some chance. To concentrate on the impact of the $\Delta\beta$ overall scaling, we normalize the noise by the largest of the 20 random numbers. All seven groups of fluctuation $\Delta\beta_1, \Delta\beta_2, ..., \Delta\beta_7$ are generated using the above method. 
}

\textcolor[rgb]{0,0,0}{
}

\subsection*{B. Experimental-designed distribution}

{The generation method is described by Eq.\ref{filter}. As the speed of writing waveguide can always be increased, but not facilitated by a decrease, the fluctuation of site energy $\Delta\beta$ is expected to be positive. For this, we take the absolute value of the noise sequence and, as with Supplementary Note 11.A, normalize them by the maximum value of the sequence.
}

{
}

\subsection*{C. Normal distribution}

{
Let a random variable $X$ have a probability density function
\begin{equation}\label{normal1}
\begin{aligned}
	p(x)=\frac{1}{\sqrt{2\pi}\sigma}e^{-\frac{{(x-\mu)}^2}{2\sigma^2}}, &&-\infty<x<\infty
\end{aligned}
\end{equation}
where $\mu,\sigma>0$ are parameters, then $X$ is said to obey a normal distribution, denoted $X\sim N(\mu,\sigma^2)$. In particular, if $\mu=0,\sigma=1$, $X$ is said to obey the standard normal distribution, denoted $X\sim N(0,1)$.
}

{
Twenty random numbers are sampled by standard normal distribution, which means $\Delta\beta\sim N(0,1)$. Then, the random numbers are taken to be positive and normalised to obtain the adopted noise sequence $\Delta\beta_i$. 
}

\subsection*{D. Gamma distribution}

{
The random variable $X$ has a probability density function
\begin{equation}\label{gamma1}
	p(x)=
\begin{cases}
\frac{\lambda^\alpha}{\Gamma(\alpha)}x^{\alpha-1}e^{-\lambda x}, & x>0,\\
0, & \rm otherwise.
\end{cases}
\end{equation}
where $\Gamma(\alpha)$ is Gamma function, then $X$ is said to obey Gamma distribution with parameters $\alpha,\lambda>0$, denoted $X\sim Ga(\alpha,\lambda)$. Specifically, if $\alpha=1$, then $X$ follows an exponential distribution with parameter $\lambda$, denoted $Exp(\lambda)$, and its probability density function is
\begin{equation}\label{exponential1}
	p(x)=
\begin{cases}
\lambda e^{-\lambda x}, & x>0,\\
0, & \rm otherwise.
\end{cases}
\end{equation}
The expectation of variable $X$ is $E(X)=\frac 1 \lambda$ for $X\sim Exp(\lambda)$. For one noise sequence, we sample 20 random numbers following exponential distribution $Exp(2)$ and normalize them. 
}

\subsection*{E. Cauchy-Lorentz distribution}

{
The random variable $X$ has a probability distribution function
\begin{equation}\label{cauchy1}
\begin{aligned}
    p(x) = \frac 1 {\pi (1+x^2)},&& -\infty<x<\infty,
\end{aligned}
\end{equation}
then $X$ is said to follow the Cauchy-Lorentz distribution.  
}

{
Suppose two mutually independent random variables $X$ and $Y$ both obey standard normal distribution $N(0,1)$. Consider a new random variable $T=\frac Y X$. The joint probability density function of $(X,Y)$ is 
\begin{equation}
p(x,y)=\varphi(x)\varphi(y),
\end{equation}
where 
\begin{equation}
    \varphi(x) = \frac 1 {\sqrt{2\pi}}e^{-\frac{x^2}{2} }
\end{equation}
is the probability density function of standard normal distribution. The probability density function of random variable $T$ is 
\begin{equation}
    \begin{aligned}
    p_T(t) &= \int_{-\infty}^\infty |x|\cdot p(x,tx)\text{d}x\\
    &= \int_{-\infty}^\infty|x|\cdot \varphi(x)\varphi(tx)\text{d}x\\
    &= \frac 1 \pi \int_0^\infty x\cdot e^{-\frac{x^2}{2}(1+t^2)}\text{d}x\\
    &=\frac{1}{\pi(1+t^2)}.
    \end{aligned}
\end{equation}
Therefore $T=\frac{Y}{X}$ obeys the Cauchy-Lorentz distribution. In other words, the quotient of two mutually independent standard normal random variables is subject to the Cauchy-Lorentz distribution.
}

{
We obtained the random numbers of the Cauchy-Lorentz distribution by calculating the quotient of two standard normal distributions. These random numbers are taken to their absolute value and normalized to give the final noise sequence $\Delta\beta$.
}

{The energy transport efficiency against the noise scale $\Delta\beta_A$ for all the different noise types above is shown in Fig.\ref{fig:Results14}. The energy transport efficiency peak always exists at a certain noise amplitude, showing the generality of the phenomenon of ENAQT. From Fig.\ref{fig:Results14}a to e, the optimal noise amplitude $\Delta\beta_A$ gradually increases and the corresponding maximum energy transport efficiency gradually decreases. This is possibly because, the noise sequences after normalization all have a maximum value of 1, but have different mean values. The trend of the best noise amplitude is in fact consistent with the mean value of the noise series. }


\newpage

\begin{figure}[ht!]
\includegraphics[width=0.95\textwidth]{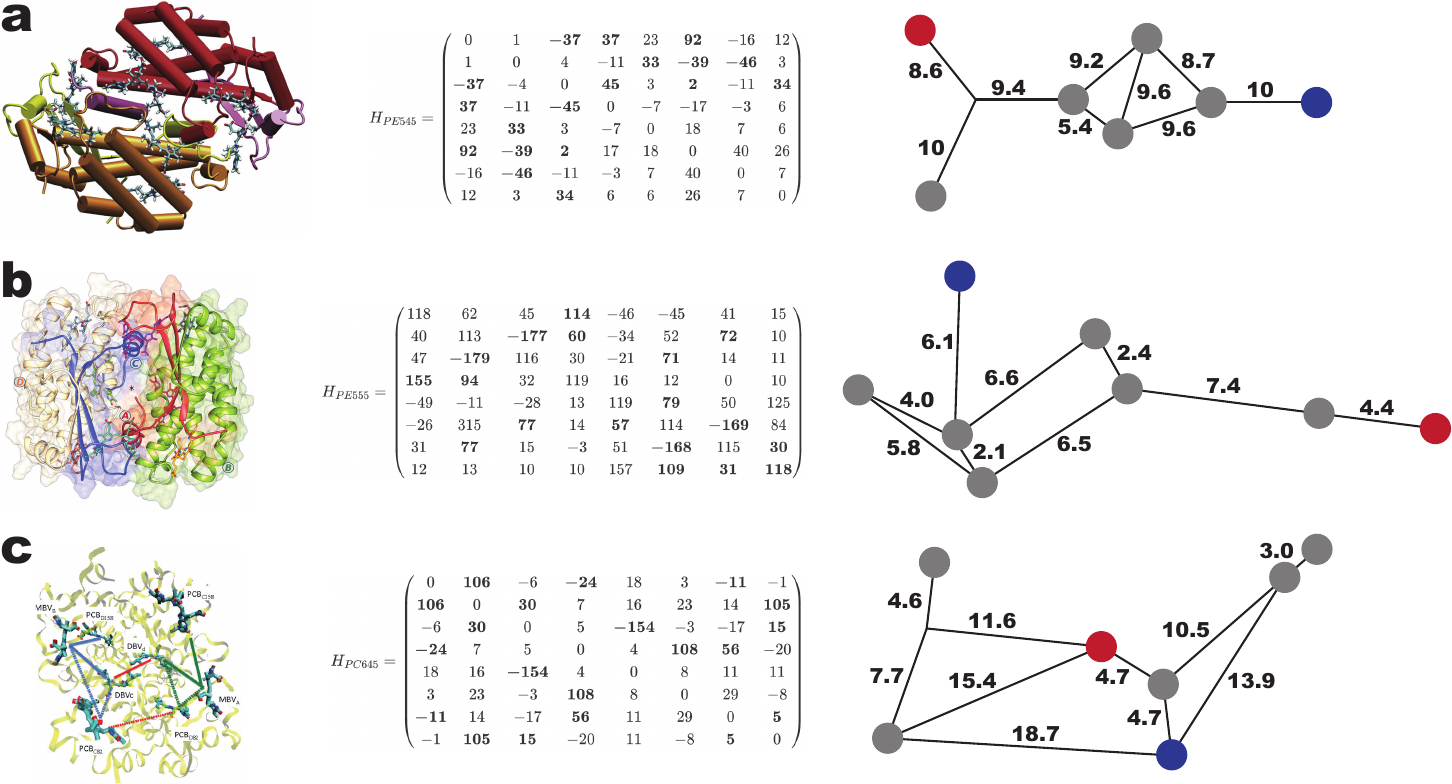}
\caption{{\textcolor[rgb]{0,0,0}{\textbf{Mapping of three different complexes on waveguides.} {\bf a,} The structure and Hamiltonian of the PE545 complex, and the waveguide layout that can mimic PE545. {\bf b,} The structure and Hamiltonian of the PE555 complex, and the waveguide layout that can mimic PE555. {\bf c,} The structure and Hamiltonian of the PE545 complex, and the waveguide layout that can mimic PC645. For all three subgraphs, the red and blue nodes respectively correspond to the input and output waveguides. The off-diagonal numbers in bold show strong coupling coefficients, which are strictly mapped on the waveguide structures. The Hamiltonians of PE545, PE555 and PC645 are referenced from Refs.\cite{Novoderezhkin2010, Chandrasekaran2016, Zech2014}, respectively. Hamiltonians are in ${\rm cm}^{-1}$. Waveguide spacings are in ${\mu\rm m}$.}}}
\label{fig:3complex}
\end{figure}

\begin{figure}[ht!]
\includegraphics[width=0.75\textwidth]{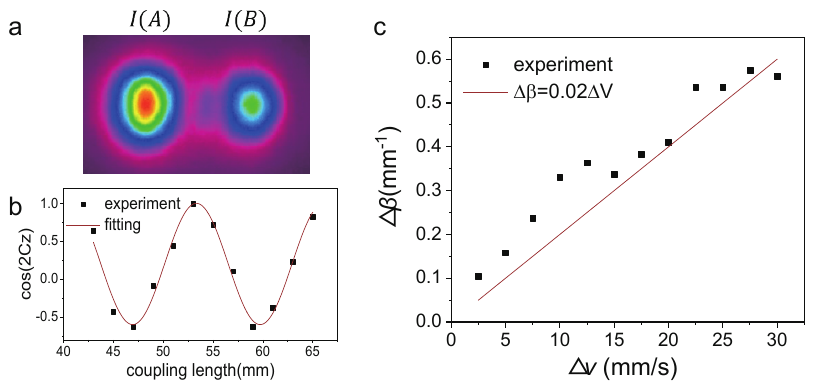}
\caption{{\textbf{The characterization of the $\Delta \beta$.} (a) The measurement of the intensity of the two modes, I(A) and I(B), of the directional coupler of a certain evolution length. (b) Fit the relationship between $\frac{I(A)-I(B)}{I(A)+I(B)}$ and $z$ with a suitable sinusoidal curve to get the coupling coefficient. (c) The experimentally measured $\Delta \beta$ against $\Delta V$.}}
\label{fig:apparato}
\end{figure}

\begin{figure*}[ht!]
\includegraphics[width=0.95\textwidth]{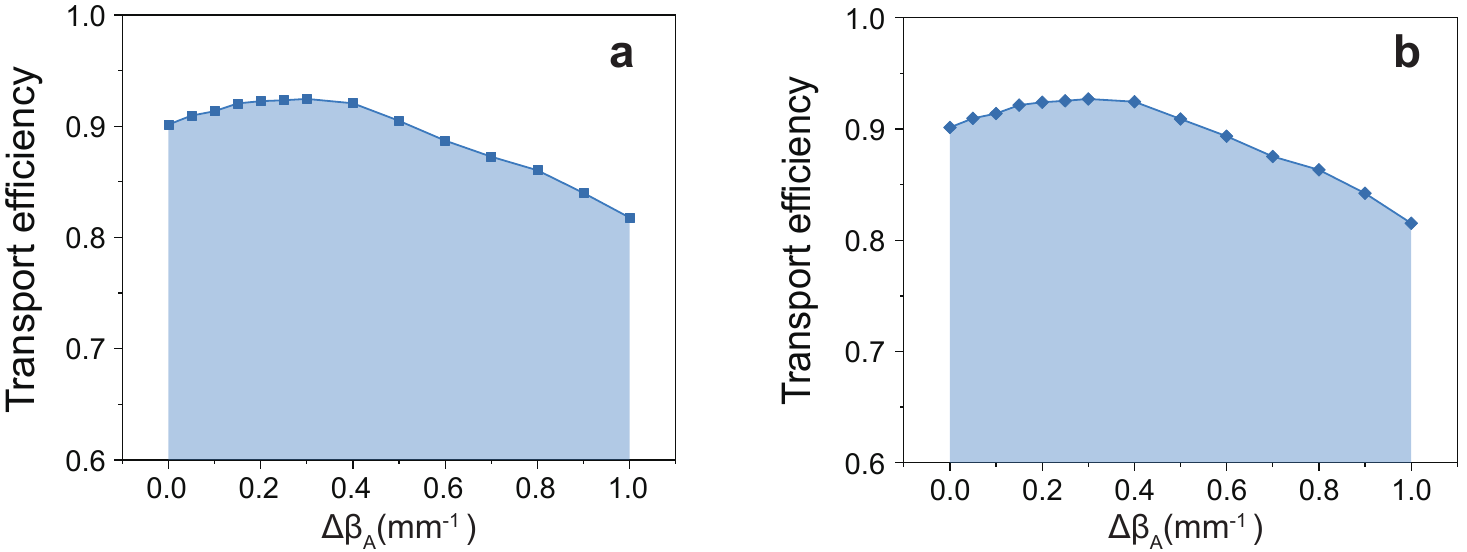}
\caption{\textbf{The calculated transport efficiency of the on-chip FMO structure under different $\Delta \beta_A$ values using the $\Delta \beta$ photonic model.} Each dot in the efficiency curve is averaged by 30 sets of random values in white noises of the same $\Delta \beta$ amplitude. In ({\bf a}), we consider both the $\Delta \beta$ and $\Delta C$ terms in the effective Hamiltonian $H_{\rm {eff}}$, while in ({\bf b}), we only consider  $\Delta \beta$ terms and ignore the $\Delta C$ terms.}
\label{fig:DeltaC}
\end{figure*}

\begin{figure*}[ht!]
\includegraphics[width=0.9\textwidth]{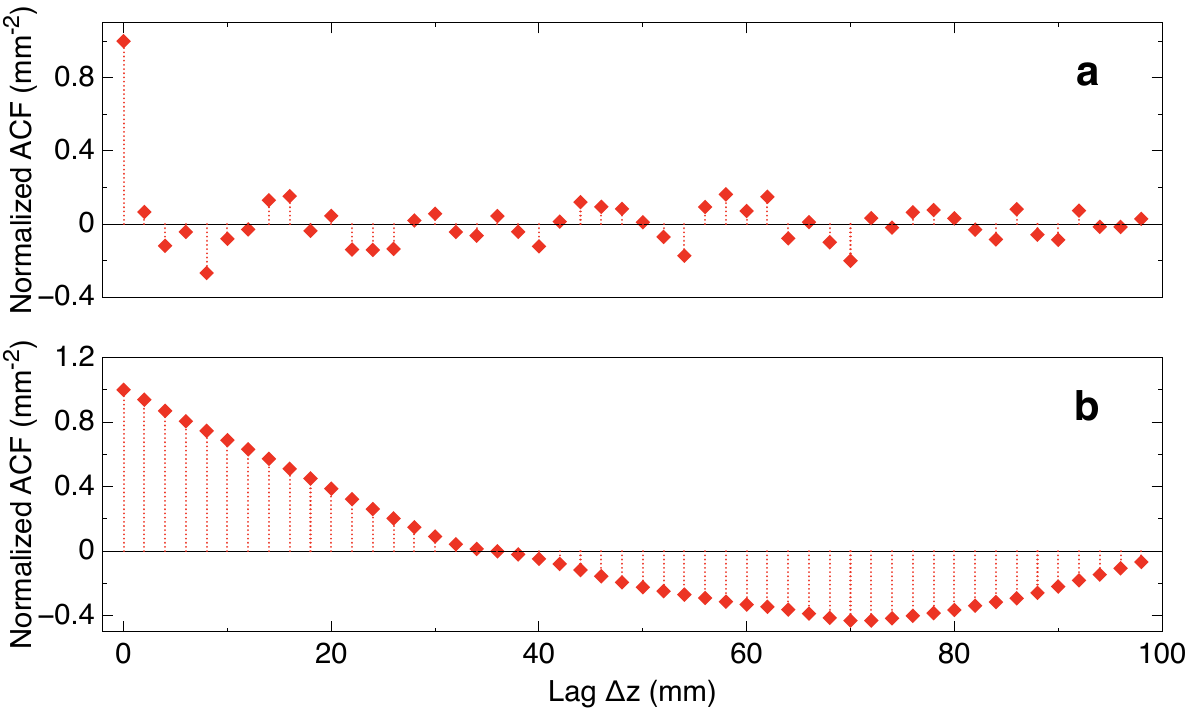}
\caption{\textbf{Normalized auto-correlation functions (ACFs) of two kinds of noise sequences}. The sampling frequency $f_s$ and sampling period $t_c$ are $0.5\ \rm mm^{-1}$ and $100\ \rm mm$, respectively. The $\Delta\beta$ amplitude $\Delta\beta_A$ is 1 $\rm mm^{-1}$. The lags of ACF are taken at the interval of 2 $\rm mm^{-1}$ in the range of 0 to 98 mm. {\bf a,} Normalized ACF of the colored noise. {\bf b,} Normalized ACF of the white noise. }
\label{fig:Results8}
\end{figure*}

\begin{figure*}[ht!]
\includegraphics[width=0.95\textwidth]{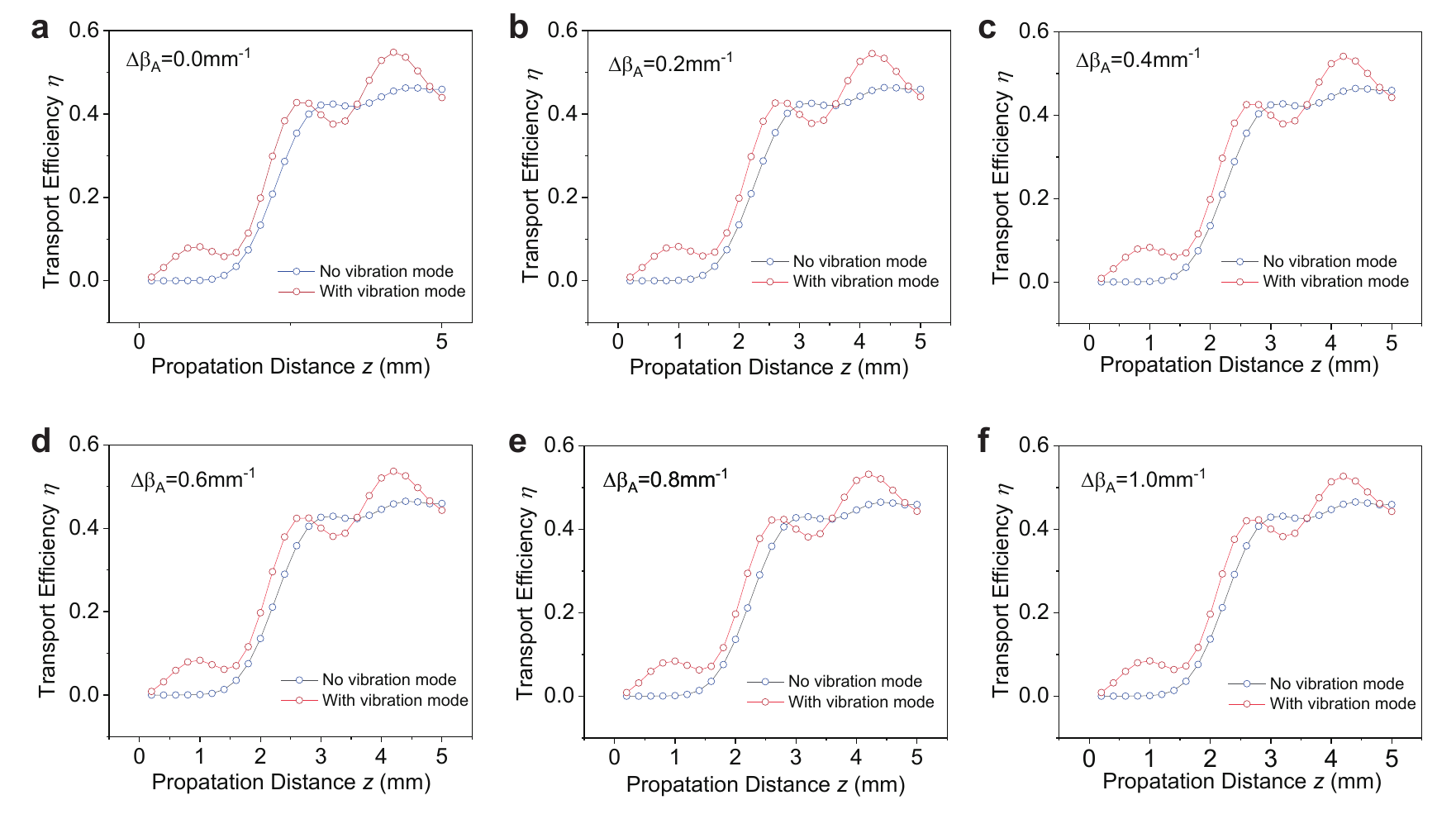}
\caption{\textbf{The vibrational assistance}. The transport efficiency at an early transport length with (red) and without (blue) vibrational assistance. The $\Delta\beta_A$ is 0, 0.2, 0.4, 0.6, 0.8, 1.0 $\rm mm^{-1}$ in {\bf a-f}, respectively.}
\label{VibAssist}
\end{figure*}

\begin{figure*}[ht!]
\includegraphics[width=\textwidth]{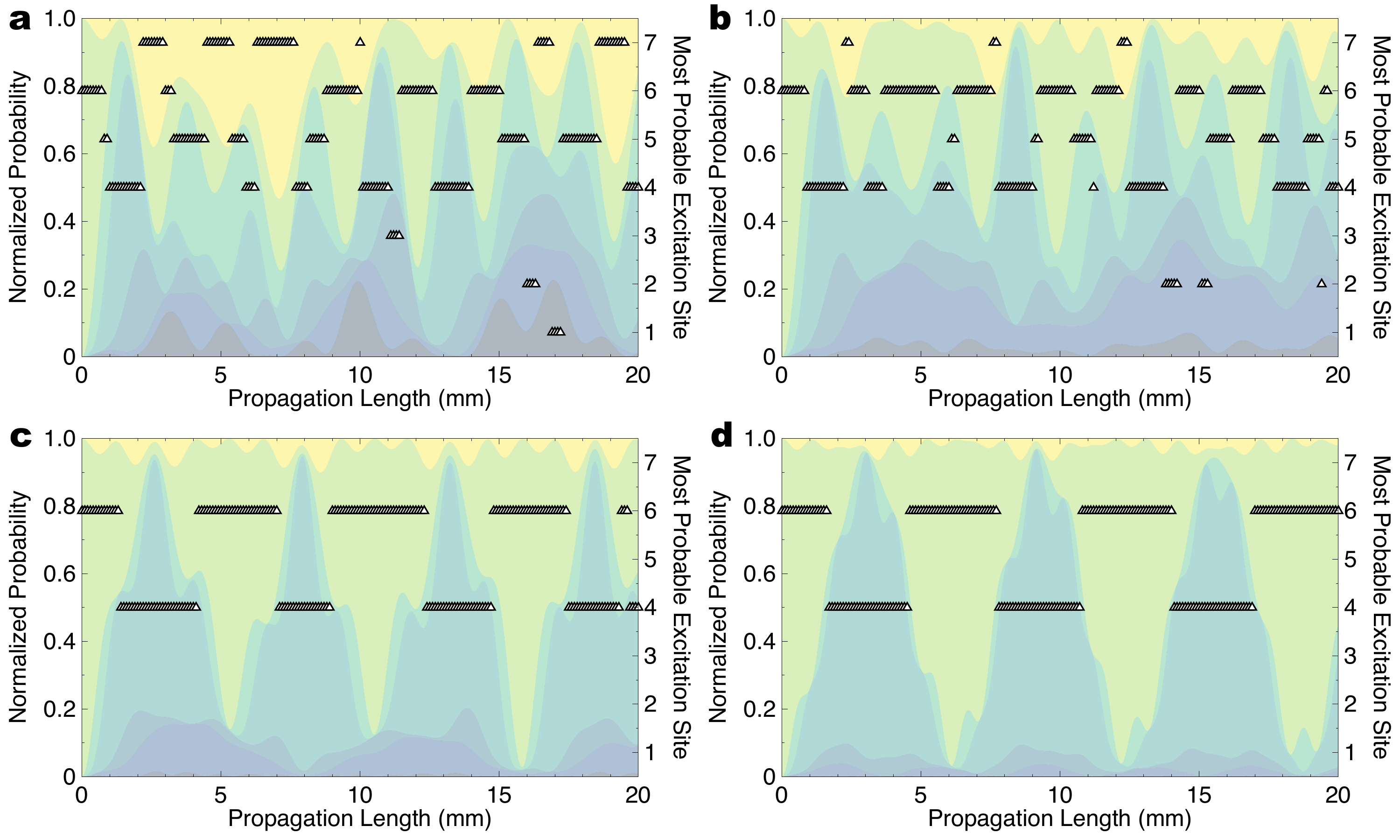}
\caption{\textbf{Normalized probability and most probable excited site with different disorder strengths}. {Different colors indicate different excitation sites. From bottom to top, different color blocks indicate the probability of excitation from Site 1 to Site 7. The sum of the normalized probabilities of different sites satisfies the normalizatin condition $\sum_{i=1}^7p_i=1$. At the initial moment, only Site 6 has an excitation probability of 1, representing laser injection only from waveguide 6. The triangles mark one of the seven sites most likely to be excited. From {\bf a} to {\bf d}, the order of disorder strength $\Gamma$ is 0, 3, 6 and 10 $\rm mm^{-1}$. All subgraphs have a noise intensity of 0.}}
\label{fig:Results17}
\end{figure*}

\begin{figure*}[ht!]
\includegraphics[width=\textwidth]{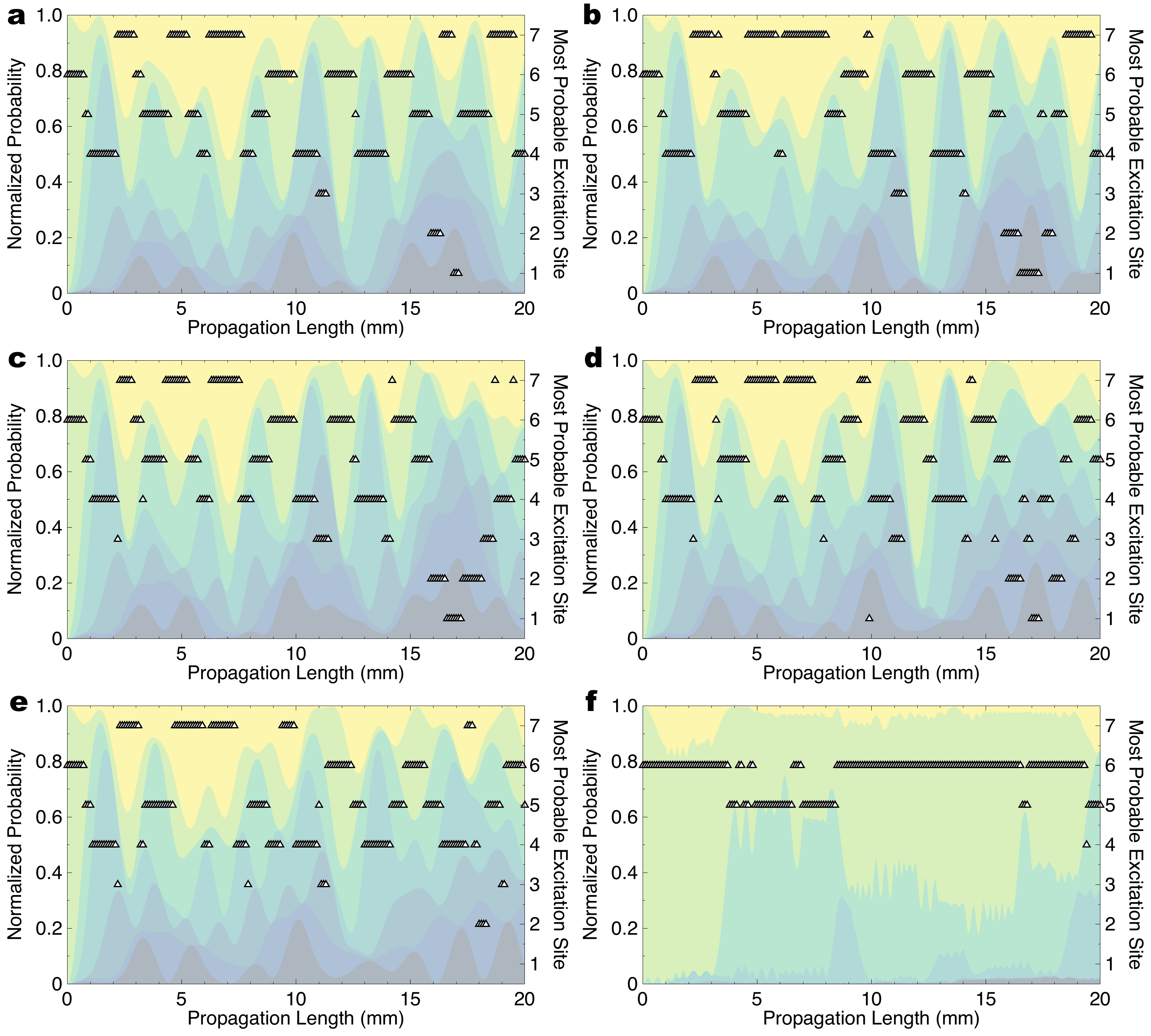}
\caption{\textbf{Normalized probability and most probable excited site with different dephasing strengths}. {Different colors indicate different excitation sites. From bottom to top, different color blocks indicate the probability of excitation from Site 1 to Site 7. The sum of the normalized probabilities of different sites satisfies the normalization condition $\sum_{i=1}^7p_i=1$. At the initial moment, only Site 6 has an excitation probability of 1, representing laser injection only from waveguide 6. The triangles mark one of the seven sites most likely to be excited. From {\bf a} to {\bf d}, the noise intensity $\Delta\beta_A$ is 0.1, 0.3, 0.5, 0.7, 1.0 and 80 $\rm mm^{-1}$. All subgraphs have a disorder intensity of 0.}}
\label{fig:Results18}
\end{figure*}

\begin{figure*}[ht!]
\includegraphics[width=\textwidth]{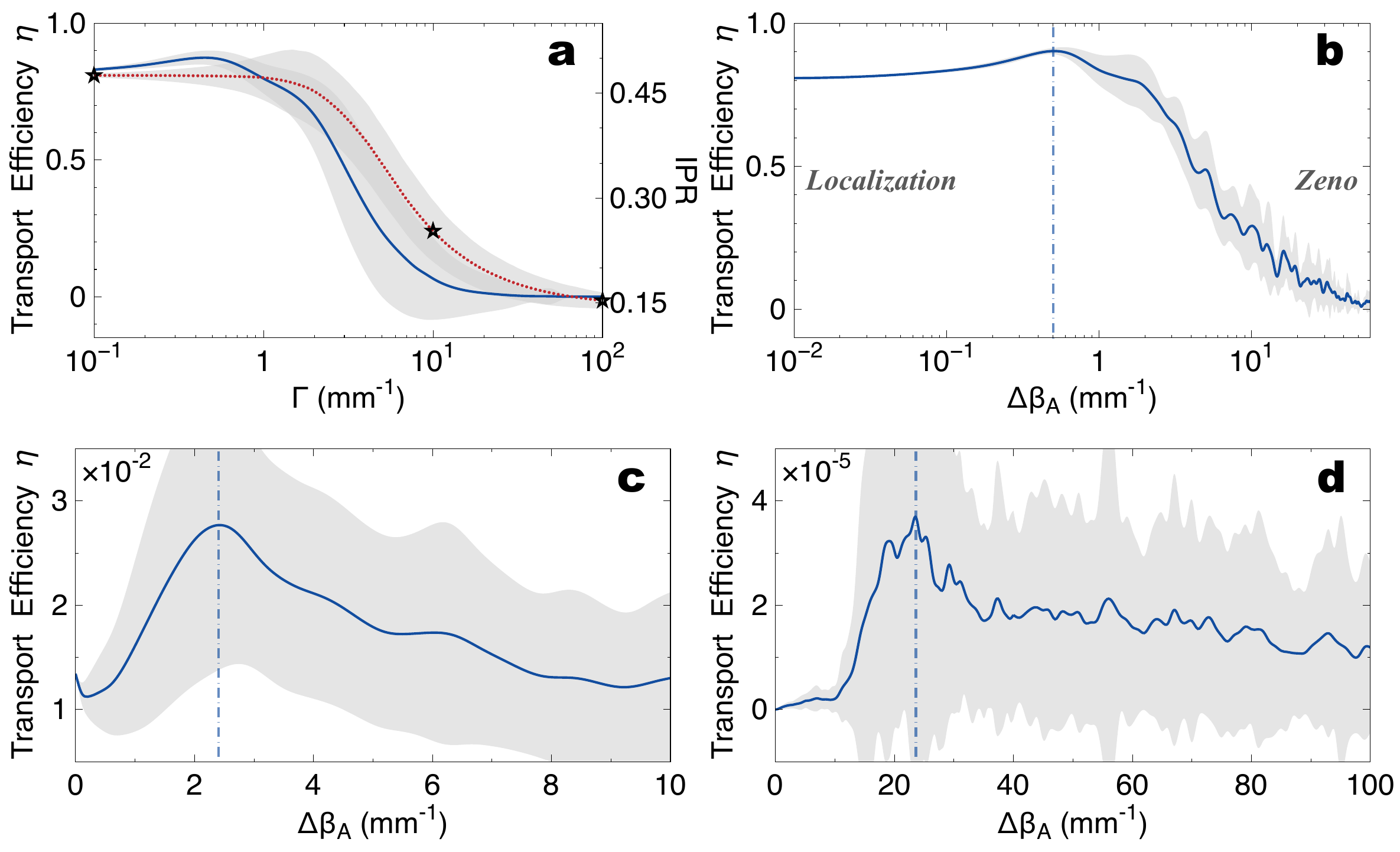}
\caption{\textbf{Energy transport efficiency with different disorder strengths $\Gamma$ and $\Delta \beta$ amplitudes}. {{\bf a, }The blue solid and red dotted line represents average energy transport efficiency and IPR at $20\rm mm$ propagation length, respectively. The shading is the standard deviation of the 1000 simulations. {\bf b}, {\bf c}, and {\bf d }are respectively simulated using different $\Delta\beta_A$ with disorder strength fixed at $\Gamma = 0$, $10$ and $100 {\rm mm^{-1}} $, as shown by the three asterisks in {\bf a}. The shading is the standard deviation of the 100 simulations. The highest average transport efficiencies present at $\Delta\beta_A=0.5\rm mm^{-1}$, $2.4\rm mm^{-1}$ and $23.6\rm mm^{-1}$. In all cases, white noise is sampled for the dephasing sequence. In the waveguide system, the transverse axis $\Delta\beta_A$ is positively correlated with the dephasing degree. Both sides of the curve demonstrate the lower transport efficiency with different suppression mechanisms, which are localization due to strong quantum coherence and Zeno effect due to strong coupling with the environment, as separately marked in {\bf b}.}}
\label{fig:Results15}
\end{figure*}

\begin{figure*}[ht!]
\includegraphics[width=\textwidth]{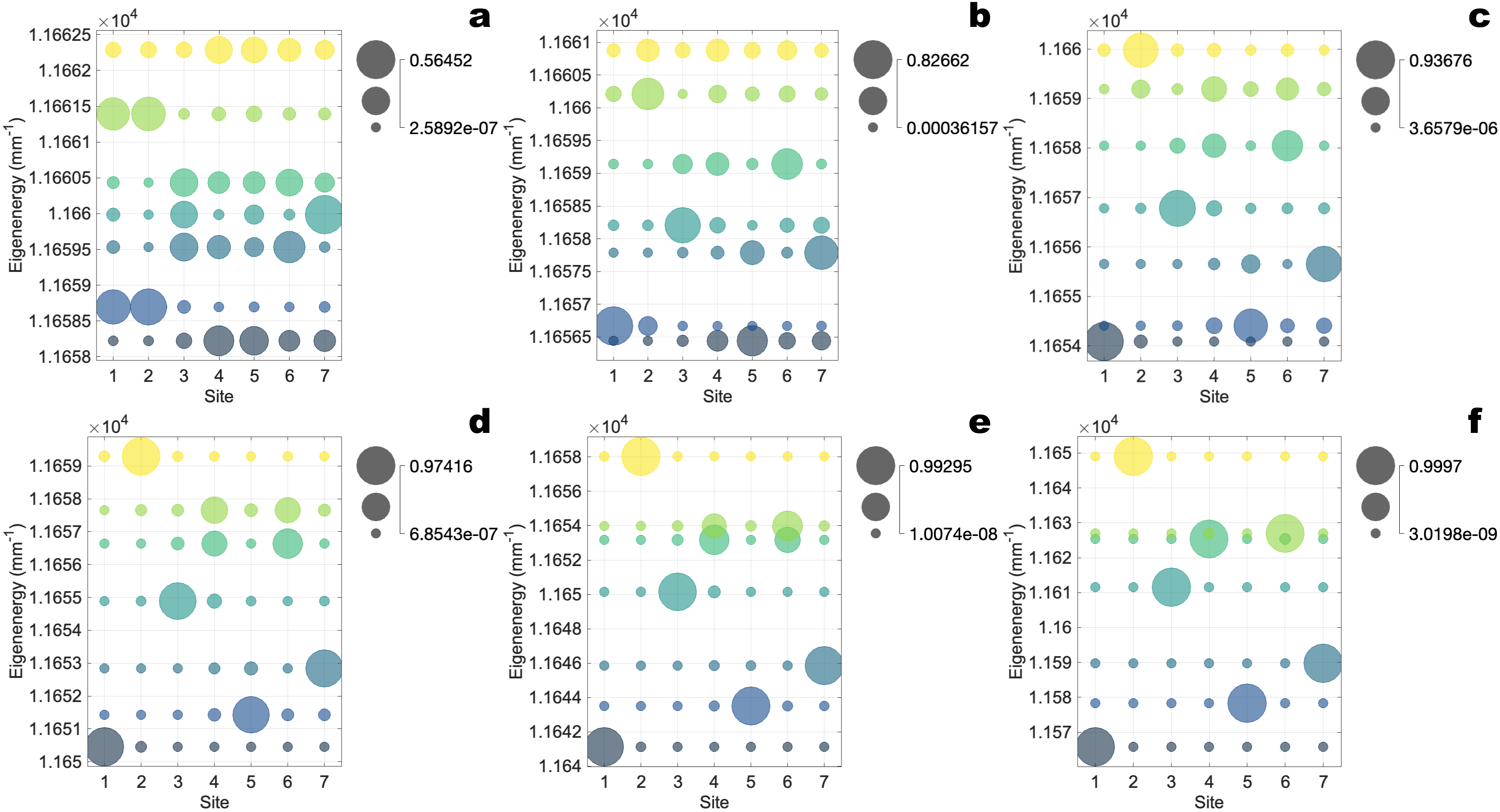}
\caption{\textbf{The distribution of eigenenergy levels under different disorders}. {The location of the bubbles indicates that the corresponding site is at a certain eigenenergy. The diameter of the bubble represents the probability of being in the specific eigenstate. For the specific correspondence between bubble diameter and probability, see the legend in the upper-right corner of each subgraph. In all cases, the system has seven different eigenenergy levels and no degenerate subspace. From {\bf a} to {\bf f}, the scale of disorder strength $\Gamma$ is 0, 3, 6, 10, 20 and 100 $\rm mm^{-1}$, respectively.}}
\label{fig:Results16}
\end{figure*}

\begin{figure*}[ht!]
\includegraphics[width=0.96\textwidth]{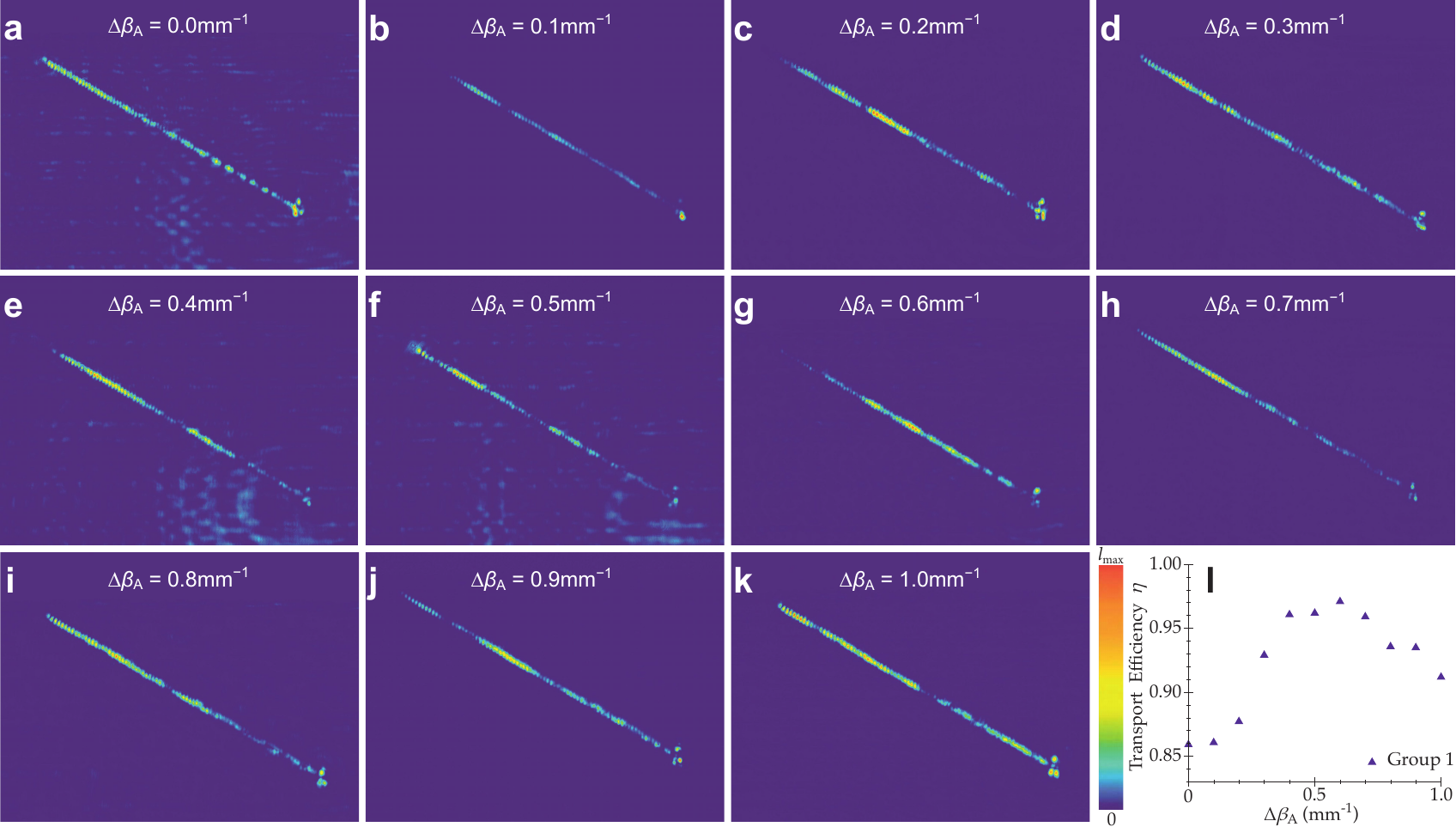}
\caption{\textbf{Experimental evolution patterns for samples of Group 1}. {\bf a-k,} Samples formed with different $\Delta \beta$ amplitudes show evolution patterns of different energy transport efficiencies. {\bf l,} The measured energy transport efficiencies against the $\Delta \beta$ amplitudes.}
\label{Group1}
\end{figure*}

\begin{figure*}[ht!]
\includegraphics[width=0.96\textwidth]{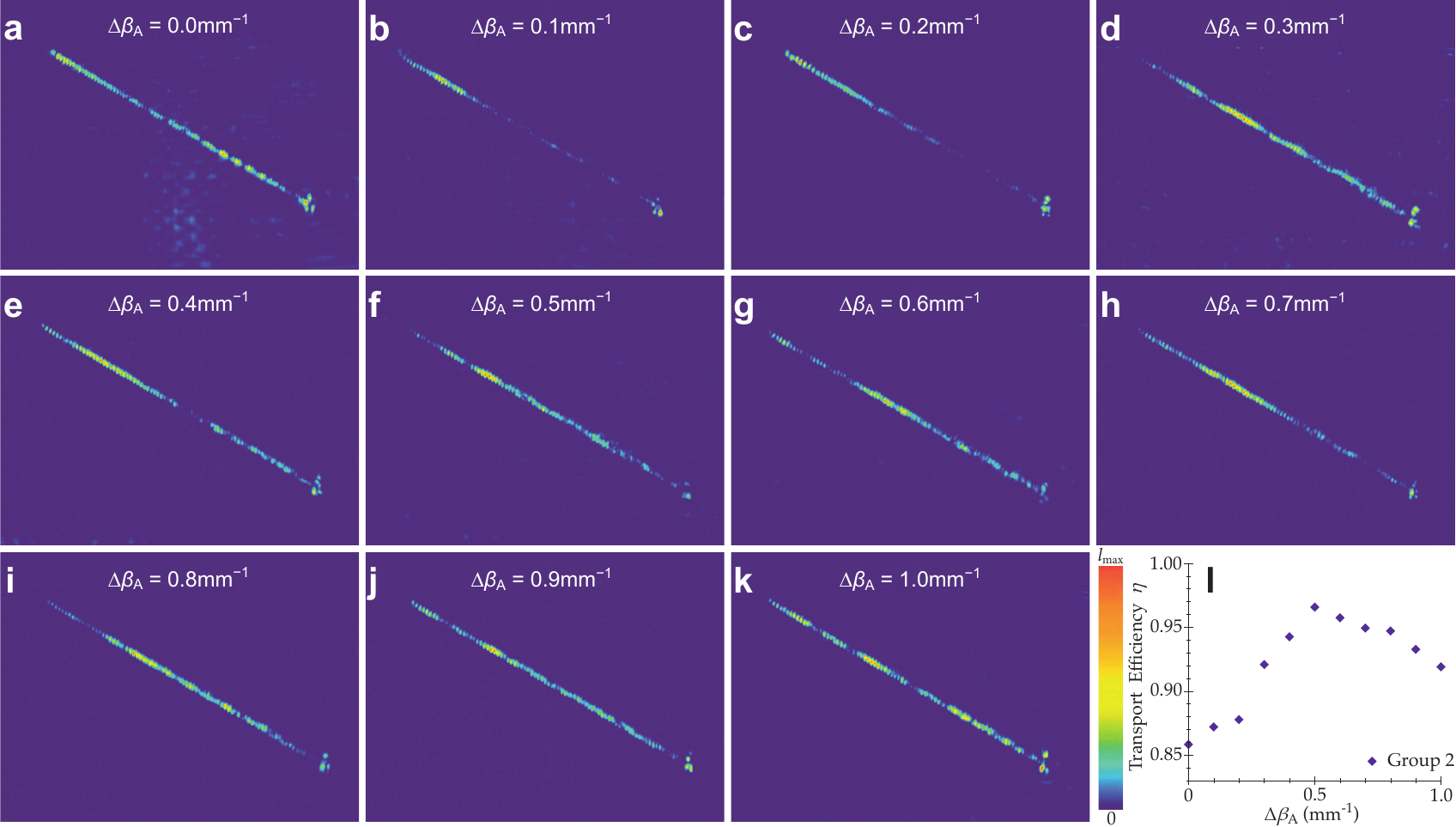}
\caption{\textbf{Experimental evolution patterns for samples of Group 2}. {\bf a-k} Samples formed with different $\Delta \beta$ amplitudes show evolution patterns of different energy transport efficiencies. {\bf l,} The measured energy transport efficiencies against the $\Delta \beta$ amplitudes.}
\label{Group2}
\end{figure*}

\begin{figure*}[ht!]
\includegraphics[width=0.96\textwidth]{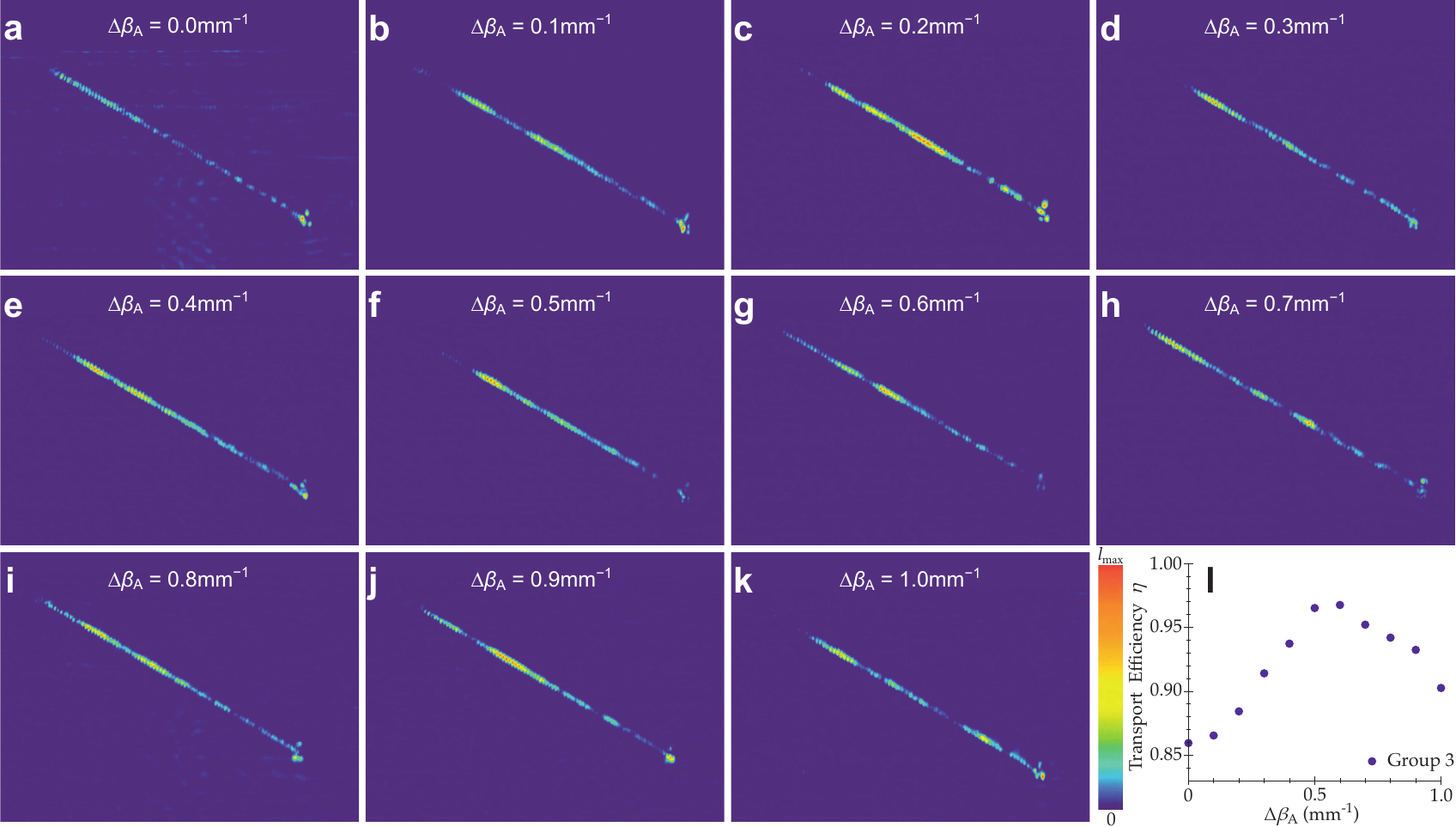}
\caption{\textbf{Experimental evolution patterns for samples of Group 3}. {\bf a-k,} Samples formed with different $\Delta \beta$ amplitudes show evolution patterns of different energy transport efficiencies. {\bf l,} The measured energy transport efficiencies against the $\Delta \beta$ amplitudes.}
\label{Group3}
\end{figure*}

\begin{figure*}[ht!]
\includegraphics[width=0.96\textwidth]{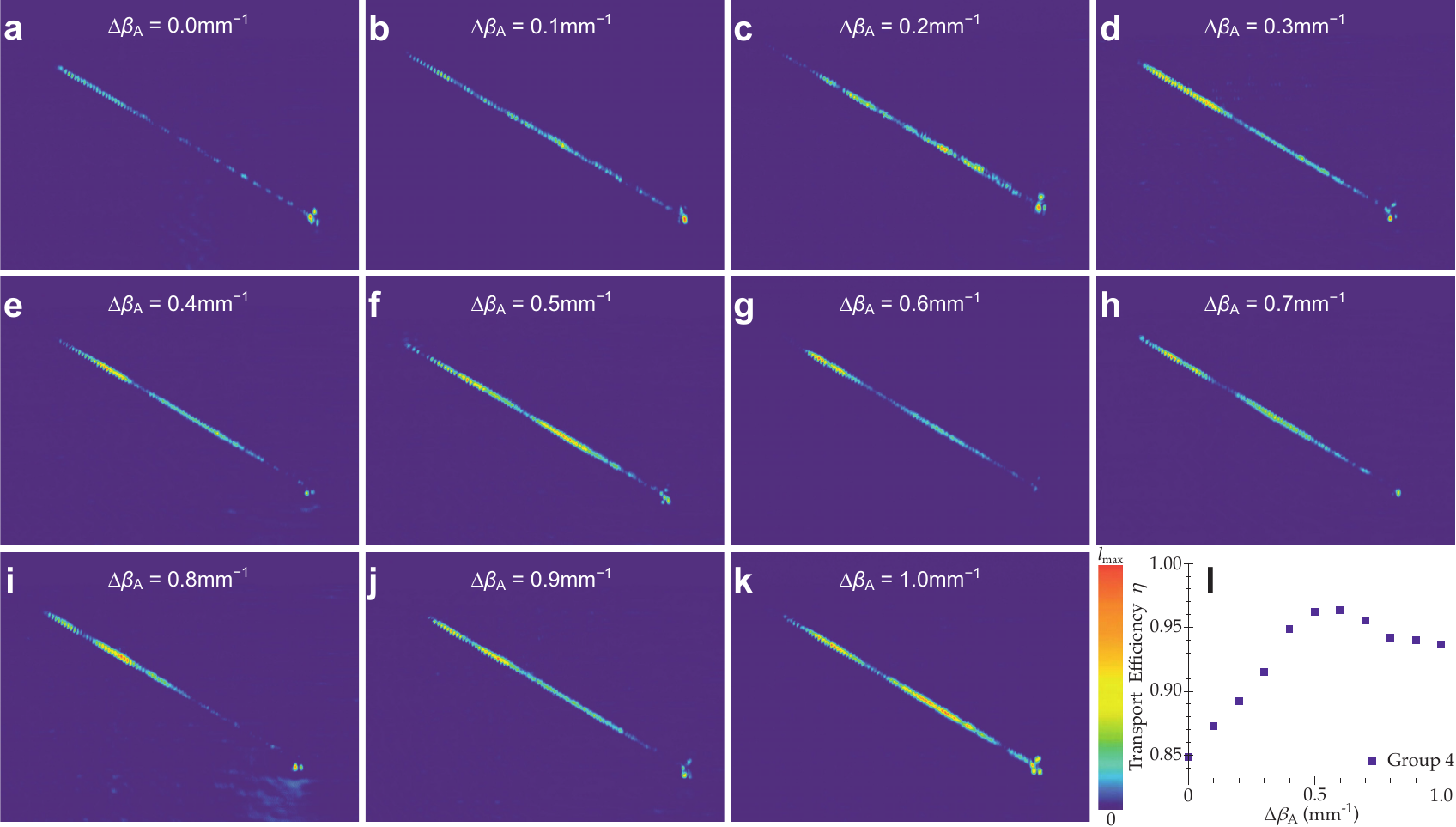}
\caption{\textbf{Experimental evolution patterns for samples of Group 4}. {\bf a-k,} Samples formed with different $\Delta \beta$ amplitudes show evolution patterns of different energy transport efficiencies. {\bf l,} The measured energy transport efficiencies against the $\Delta \beta$ amplitudes.}
\label{Group4}
\end{figure*}

\begin{figure*}[ht!]
\includegraphics[width=0.96\textwidth]{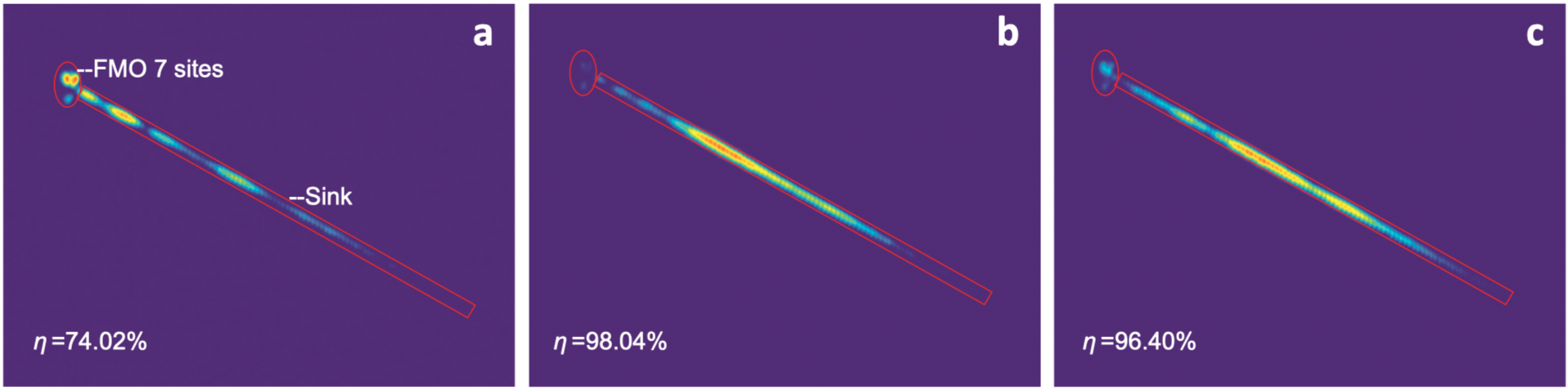}
\caption{\textcolor[rgb]{0,0,0}{\textbf{Experimental transport patterns of white-noise FMO waveguides.} Three examples from the arrays formed with different $\Delta\beta_A$ values show different energy transport efficiencies. The $\Delta\beta_A$ values are $0.0{\rm mm}^{-1}$ for {\bf a}, $0.4{\rm mm}^{-1}$ for {\bf b} and $0.9{\rm mm}^{-1}$ for {\bf c}. The zones for seven-site FMO complex and the sink are marked with ellipses and rectangles respectively. The energy transport efficiencies are given in the bottom-left of the figures.}}
\label{FigureS21}
\end{figure*}

\begin{figure*}[ht!]
\includegraphics[width=0.6\textwidth]{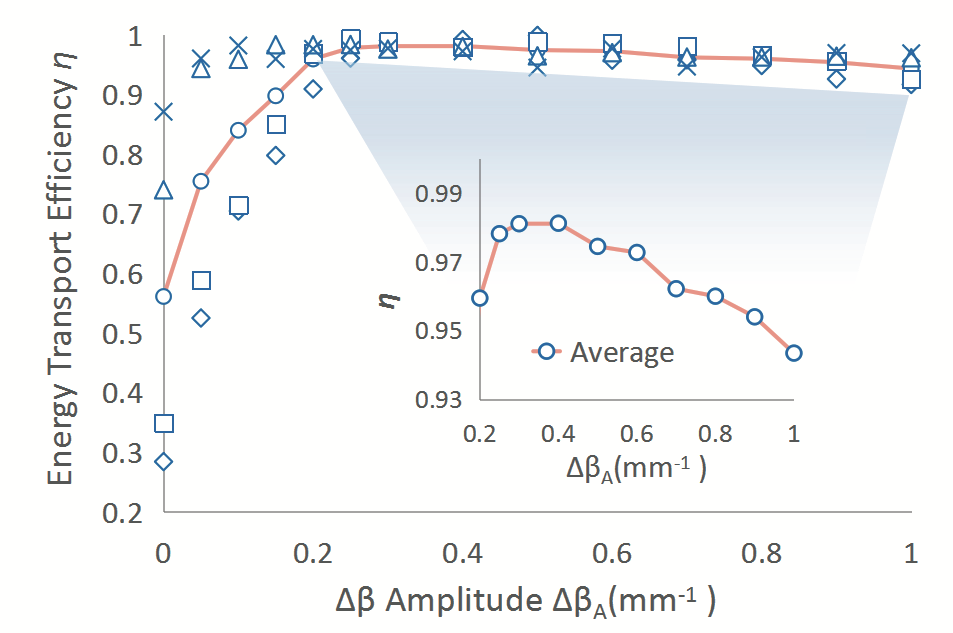}
\caption{\textcolor[rgb]{0,0,0}{\textbf{Experimental energy transport efficiency.} The measured transport efficiencies for samples of different $\Delta\beta_A$ values. The results for each individual sample and the averaged values are plotted in dots and a curve, respectively. Inset shows part of the energy transport efficiency curve where the droop is prominent.}}
\label{FigureS20}
\end{figure*}

\begin{figure*}[ht!]
\includegraphics[width=\textwidth]{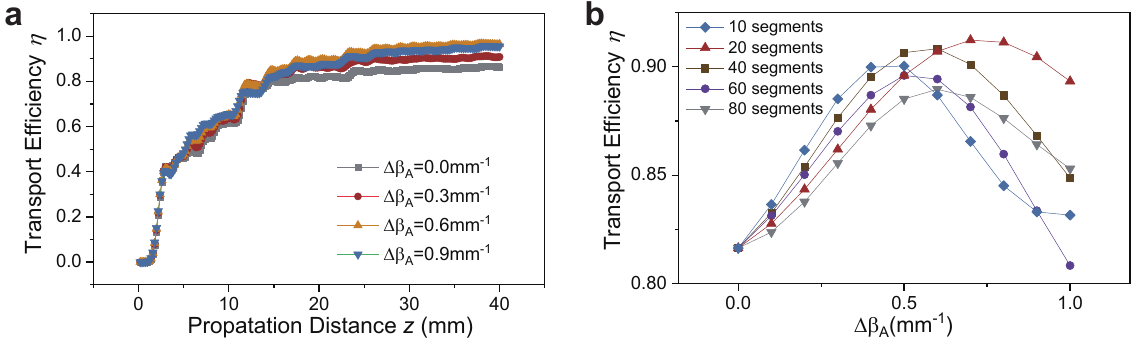}
\caption{\textbf{Simulated energy transport efficiency}. {\bf a,} The energy transport efficiencies against the propagation distance under various $\Delta\beta$ amplitudes, namely, 0, 0.3 $\rm mm^{-1}$, 0.6 $\rm mm^{-1}$ and 0.9 $\rm mm^{-1}$. The segment length for these samples is set to 1mm. {\bf b,} The energy transport efficiency at a propagation distance of 20mm for different settings of the segment length, which set 20mm into 10, 20, 40, 60, or 80 segments, respectively.}
\label{fig:Results6}
\end{figure*}

\begin{figure*}[ht!]
\includegraphics[width=\textwidth]{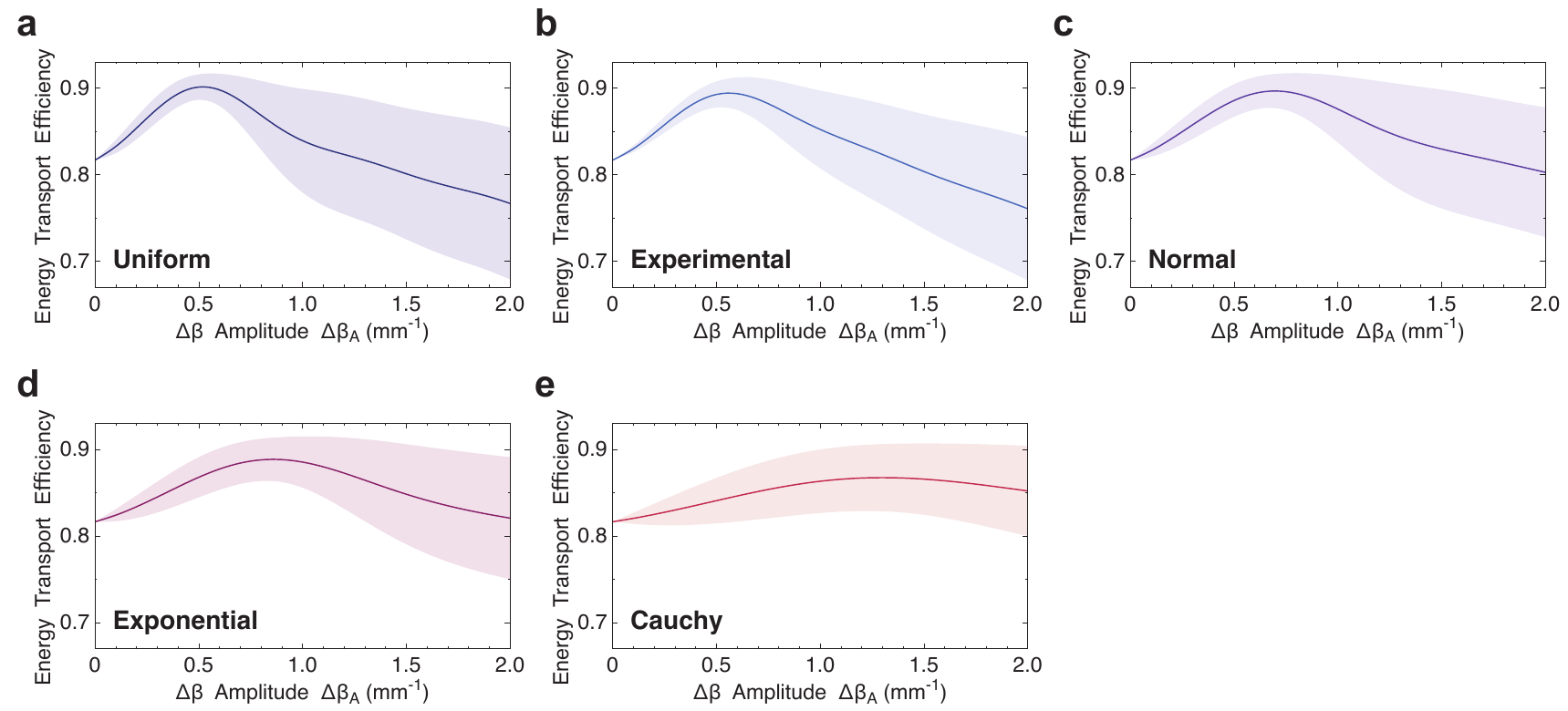}
\caption{\textbf{Energy transport efficiency with different noise distributions}. {The type of probability density function of the noise distribution is shown in the bottom left of the figure. }}
\label{fig:Results14}
\end{figure*}


\begin{thebibliography}{10}
\renewcommand{\bibnumfmt}[1]{#1.}



\bibitem{Engel2007}
\bibinfo{author} {Engel, G. S., Calhoun, T. R., Read, E. L., Ahn, T. K., Man\v{c}al, T., Cheng, Y. C., Blankenship, R. E., \& Fleming, G. R.} 
\bibinfo{title} {Evidence for wavelike energy transfer through quantum coherence in photosynthetic systems.}
\newblock \emph{\bibinfo{journal}{Nature}}
\textbf{\bibinfo{volume}{446}}, \bibinfo{pages}{782-786}
  (\bibinfo{year}{2007}).


\bibitem{Lee2007}
\bibinfo{author} {Lee, H., Cheng, Y. C., \& Fleming, G. R.} 
\bibinfo{title}{Coherence dynamics in photosynthesis: protein protection of excitonic coherence.}
\newblock \emph{\bibinfo{journal}{Science}}
\textbf{\bibinfo{volume}{316}}, \bibinfo{pages}{1462-1465}
  (\bibinfo{year}{2007}).

\bibitem{Collini2010}    
\bibinfo{author} {Collini, E., Wong, C. Y., Wilk, K. E., Curmi, P. M., Brumer, P., \& Scholes, G. D.} 
\bibinfo{title}{Coherently wired light-harvesting in photosynthetic marine algae at ambient temperature.}
\newblock \emph{\bibinfo{journal}{Nature}}
\textbf{\bibinfo{volume}{463}}, \bibinfo{pages}{644}
  (\bibinfo{year}{2010}).

\bibitem{Panitch2010}
\bibinfo{author} {Panitchayangkoon, G., Hayes, D., Fransted, K. A., Caram, J. R., Harel, E., Wen, J., Blankenship, R. E., \& Engel, G. S.} 
\bibinfo{title}{Long-lived quantum coherence in photosynthetic complexes at physiological temperature.}
\newblock \emph{\bibinfo{journal}{Proc. Nat. Aca. Sci.}}
\textbf{\bibinfo{volume}{107}}, \bibinfo{pages}{12766-12770}
  (\bibinfo{year}{2010}).


\bibitem{Breuer2007}
\bibinfo{author} {Breuer, H.-P. \& Petruccione, F.} 
\newblock \emph{\bibinfo{journal}{The Theory of Open Quantum Systems.}}
\bibinfo{pages}{Oxford University Press}
  (\bibinfo{year}{2007}).

\bibitem{Mohseni2008}
\bibinfo{author} {Mohseni, M., Robentrost, P., Lloyd, S. \& Aspuru-Guzik, A.} 
\bibinfo{title} {Environment-assisted quantum walks in photosynthetic energy transfer.}
\newblock \emph{\bibinfo{journal}{ J. Chem. Phys.}}
\textbf{\bibinfo{volume}{129}}, \bibinfo{pages}{176106}
  (\bibinfo{year}{2008}).


\bibitem{Plenio2008}
\bibinfo{author} {Plenio, M. B., \& Huelga, S. F.} 
\bibinfo{title} {Dephasing-assisted transport: quantum networks and biomolecules.}
\newblock \emph{\bibinfo{journal}{New J. Phys.}}
\textbf{\bibinfo{volume}{10}}, \bibinfo{pages}{113019}
  (\bibinfo{year}{2008}).

\bibitem{Caruso2009}
\bibinfo{author} {Caruso, F., Chin, A. W., Datta, A., Huelga, S. F., \& Plenio, M. B.} 
\bibinfo{title} {Highly efficient energy excitation transfer in light-harvesting complexes: The fundamental role of noise-assisted transport.}
\newblock \emph{\bibinfo{journal}{J. Chem. Phys.}}
\textbf{\bibinfo{volume}{131}}, \bibinfo{pages}{09B612}
  (\bibinfo{year}{2009}).

\bibitem{Caruso2014}
\bibinfo{author} {Caruso, F.} 
\bibinfo{title} {Universally optimal noisy quantum walks on complex networks.}
\newblock \emph{\bibinfo{journal}{New J. Phys.}}
\textbf{\bibinfo{volume}{16}}, \bibinfo{pages}{055015}
(\bibinfo{year}{2014}).

\bibitem{Wu2010}
\bibinfo{author} {Wu, J., Liu, F., Shen, Y., Cao, J., \& Silbey, R. J.} 
\bibinfo{title} {Efficient energy transfer in light-harvesting systems, I: optimal temperature, reorganization energy and spatial–temporal correlations.}
\newblock \emph{\bibinfo{journal}{New J. Phys.}}
\textbf{\bibinfo{volume}{12}}, \bibinfo{pages}{105012}
  (\bibinfo{year}{2010}).


\bibitem{Rebentrost2009}
\bibinfo{author} {Rebentrost, P., Mohseni, M., Kassal, I., Lloyd, S. \& Aspuru-Guzik, A.} 
\bibinfo{title} {Environment-assisted quantum transport.}
\newblock \emph{\bibinfo{journal}{New J. Phys.}}
\textbf{\bibinfo{volume}{11}}, \bibinfo{pages}{033003}
  (\bibinfo{year}{2009}).

\bibitem{Tao2020a}
\bibinfo{author} {Tao, M. J., Zhang, N. N., Wen, P. Y., et al.} 
\bibinfo{title} {Coherent and incoherent theories for photosynthetic energy transfer.}
\newblock \emph{\bibinfo{journal}{Sci. Bull.}}
\textbf{\bibinfo{volume}{65}}, \bibinfo{pages}{318-328}
  (\bibinfo{year}{2020}).

\bibitem{Park2016}
\bibinfo{author} {Park, H., Heldman, N., Rebentrost, P., et al.} 
\bibinfo{title} {Enhanced energy transport in genetically engineered excitonic networks.}
\newblock \emph{\bibinfo{journal}{Nat. Mater.}}
\textbf{\bibinfo{volume}{15}}, \bibinfo{pages}{211-216}
  (\bibinfo{year}{2016}).

\bibitem{Scholes2011}
\bibinfo{author} {Scholes, G. D., Fleming, G. R., Olaya-Castro, A., \& Van Grondelle, R.} 
\bibinfo{title}{Lessons from nature about solar light harvesting.}
\newblock \emph{\bibinfo{journal}{Nat. Chem.}}
\textbf{\bibinfo{volume}{3}}, \bibinfo{pages}{763}
  (\bibinfo{year}{2011}).

\bibitem{Lambert2013}
\bibinfo{author}{Lambert, N., Chen, Y. N., Cheng, Y. C., Li, C. M., Chen, G. Y., \& Nori, F. }
\newblock \bibinfo{title}{Quantum biology.}
\newblock \emph{\bibinfo{journal}{Nat. Phys.}}
\textbf{\bibinfo{volume}{9}}, \bibinfo{pages}{10-18} 
(\bibinfo{year}{2013}).

\bibitem{Fenna1975}
\bibinfo{author} {Fenna, R. E., \& Matthews, B. W.} 
\bibinfo{title} {Chlorophyll arrangement in a bacteriochlorophyll protein from \emph{Chlorobium limicola}.}
\newblock \emph{\bibinfo{journal}{Nature}}
\textbf{\bibinfo{volume}{258}}, \bibinfo{pages}{573-577}
  (\bibinfo{year}{1975}).

\bibitem{Hase2017}
\bibinfo{author} {H\"ase, F., Kreisbeck, C., \& Aspuru-Guzik, A.} 
\bibinfo{title} {Machine learning for quantum dynamics: deep learning of excitation energy transfer properties.}
\newblock \emph{\bibinfo{journal}{Chem. Sci.}}
\textbf{\bibinfo{volume}{8}}, \bibinfo{pages}{8419-8426}
  (\bibinfo{year}{2017}).

\bibitem{Biggerstaff2016}    
\bibinfo{author} {Biggerstaff, D.N., Heilmann, R., Zecevik, A.A., Gr\"afe, M., Broome, M.A., Fedrizzi, A., Nolte, S., Szameit, A., White, A.G., \&Kassal, I.} 
\bibinfo{title} {Enhancing quantum transport in a photonic network using controllable decoherence.}
\newblock \emph{\bibinfo{journal}{Nat. Commun.}}
\textbf{\bibinfo{volume}{7}}, \bibinfo{pages}{11282}
(\bibinfo{year}{2016}).

\bibitem{Potocnik2018}
\bibinfo{author}{Poto\v{c}nik, A., Bargerbos, A., Schr\"oder, F.A., Khan, S.A., Collodo, M.C., Gasparinetti, S., Salath\'e, Y., Creatore, C., Eichler, C., T\"ureci, H.E., \& Chin, A.W.}
\newblock \bibinfo{title}{Studying light-harvesting models with superconducting circuits.}
\newblock \emph{\bibinfo{journal}{Nat. Commun.}}
\textbf{\bibinfo{volume}{9}}, \bibinfo{pages}{904} 
(\bibinfo{year}{2018}).

\bibitem{Wang2018}
\bibinfo{author}{Wang, B. X., Tao, M. J., Ai, Q., Xin, T., Lambert, N. W., Ruan, D., Cheng, Y. C., Nori, F., Deng, F. G., \& Long, G. L.}
\newblock \bibinfo{title}{Efficient quantum simulation of photosynthetic light harvesting.}
\newblock \emph{\bibinfo{journal}{npj Quantum Info.}}
\textbf{\bibinfo{volume}{4}}, \bibinfo{pages}{52} 
 (\bibinfo{year}{2018}).

 \bibitem{Tao2020b}
\bibinfo{author} {Tao, M. J., Hua, M., Zhang, N. N., et al.} 
\bibinfo{title} {Quantum simulation of clustered photosynthetic light harvesting in a superconducting quantum circuit.}
\newblock \emph{\bibinfo{journal}{Quantum Engineering}}
\textbf{\bibinfo{volume}{2}}, \bibinfo{pages}{e53}
  (\bibinfo{year}{2020}).

\bibitem{Harris2017}
\bibinfo{author}{Harris, N.C., Steinbrecher, G.R., Prabhu, M., Lahini, Y., Mower, J., Bunandar, D., Chen, C., Wong, F. N.C., Baehr-Jones, T., Hochberg, M., Lloyd, S., \& Englund, D. }
\newblock \bibinfo{title}{Quantum transport simulations in a programmable nanophotonic processor.}
\newblock \emph{\bibinfo{journal}{Nat. Photon.}}
\textbf{\bibinfo{volume}{11}}, \bibinfo{pages}{447-452}
  (\bibinfo{year}{2017}).

\bibitem{Maier2019}
\bibinfo{author} {Maier, C., Brydges, T., Jurcevic, P., Trautmann, N., Hempel, C., Lanyon, B. P., Hauke, P., Blatt, R., \& Roos, C. F.} 
\bibinfo{title} {Environment-Assisted Quantum Transport in a 10-qubit Network.}
\newblock \emph{\bibinfo{journal}{Phys. Rev. Lett.}}
\textbf{\bibinfo{volume}{122}}, \bibinfo{pages}{050501}
  (\bibinfo{year}{2019}).

\bibitem{Tang2020}
\bibinfo{author} {Tang, H., Shi, R. X., He, T. S., Zhu, Y. Y., Wang, T. Y., Lee, M., \& Jin, X. M.} 
\bibinfo{title} {TensorFlow solver for quantum PageRank in large-scale networks.}
\newblock \emph{\bibinfo{journal}{Sci. Bull.}}
\textbf{\bibinfo{volume}{66}}, \bibinfo{pages}{120-126}
  (\bibinfo{year}{2021}).


\bibitem{Cao2020}
\bibinfo{author} {Cao, J., Cogdell, R.J., Coker, D.F., Duan, H.G., Hauer, J., Kleinekathöfer, U., $et al.$} 
\bibinfo{title} {Quantum biology revisited.}
\newblock \emph{\bibinfo{journal}{Sci. Adv.}}
\textbf{\bibinfo{volume}{6}}, \bibinfo{pages}{eaaz4888}
  (\bibinfo{year}{2020}). 
  
\bibitem{Thyrhaug2018}
\bibinfo{author} {Thyrhaug, E., Tempelaar, R., Alcocer, M.J., \v Z\'idek, K., Bína, D., Knoester, J., Jansen, T.L. \& Zigmantas, D.} 
\bibinfo{title} {Identification and characterization of diverse coherences in the Fenna–Matthews–Olson complex.}
\newblock \emph{\bibinfo{journal}{Nat. Chem.}}
\textbf{\bibinfo{volume}{10}}, \bibinfo{pages}{780-786}
  (\bibinfo{year}{2018}). 
  
\bibitem{Mancal2020}
\bibinfo{author} {Man\v cal, T.} 
\bibinfo{title} {A decade with quantum coherence: How our past became classical and the future turned quantum.}
\newblock \emph{\bibinfo{journal}{Chem. Phys.}}
\textbf{\bibinfo{volume}{532}}, \bibinfo{pages}{110663}
  (\bibinfo{year}{2020}). 
  

\bibitem{Ishizaki2021}
\bibinfo{author} {Ishizaki, A. \& Fleming, G. R.} 
\bibinfo{title} {Insights into Photosynthetic Energy Transfer Gained from Free-Energy Structure: Coherent Transport, Incoherent Hopping, and Vibrational Assistance Revisited.}
\newblock \emph{\bibinfo{journal}{J. Phys. Chem. B}}
\textbf{\bibinfo{volume}{125}}, \bibinfo{pages}{3286-3295}
  (\bibinfo{year}{2021}). 

\bibitem{Wendling2000}
\bibinfo{author} {Wendling, M., Pullerits, T., Przyjalgowski, M.A., Vulto, S.I., Aartsma, T.J., van Grondelle, R. \& van Amerongen, H.} 
\bibinfo{title} {Electron-Vibrational Coupling in the Fenna-Matthews-Olson Complex of Prosthecochloris a estuarii Determined by Temperature-Dependent Absorption and Fluorescence Line-Narrowing Measurements.}
\newblock \emph{\bibinfo{journal}{J. Phys. Chem. B}}
\textbf{\bibinfo{volume}{104}}, \bibinfo{pages}{5825-5831}
  (\bibinfo{year}{2000}). 

\bibitem{Klinger2020}
\bibinfo{author} {Klinger, A., Lindorfer, D., M\"uh, F., \&Renger, T.} 
\bibinfo{title} {Normal mode analysis of spectral density
of FMO trimers: Intra- and intermonomer energy transfer.}
\newblock \emph{\bibinfo{journal}{J. Chem. Phys.}}
\textbf{\bibinfo{volume}{153}}, \bibinfo{pages}{215103}
  (\bibinfo{year}{2020}). 

\bibitem{Ishizaki2009}
\bibinfo{author} {Ishizaki, A. \&Fleming, G. R.} 
\bibinfo{title} {Unified treatment of quantum coherent and incoherent hopping dynamics in electronic energy transfer: Reduced hierarchy equation approach.}
\newblock \emph{\bibinfo{journal}{J. Chem. Phys.}}
\textbf{\bibinfo{volume}{130}}, \bibinfo{pages}{234111}
  (\bibinfo{year}{2009}). 
  
\bibitem{Jang2008}
\bibinfo{author} {Jang, S., Cheng, Y. C., Reichman, D. R., \& Eaves, J. D.} 
\bibinfo{title} {Theory of coherent resonance energy transfer.}
\newblock \emph{\bibinfo{journal}{J. Chem. Phys.}}
\textbf{\bibinfo{volume}{129}}, \bibinfo{pages}{101104}
  (\bibinfo{year}{2008}). 

\bibitem{Chen2022}
\bibinfo{author} {Chen, X. Y., Zhang, N. N., He, W. T., Kong, X. Y., Tao, M. J., Deng, F. G., Ai, Q., \& Long, G. L.} 
\bibinfo{title} {Global correlation and local information flows in controllable non-Markovian open quantum dynamics.}
\newblock \emph{\bibinfo{journal}{npj Quantum Info.}}
\textbf{\bibinfo{volume}{8}}, \bibinfo{pages}{1-6}
  (\bibinfo{year}{2022}). 


\bibitem{Caruso2016}
\bibinfo{author} {Caruso, F., Crespi, A., Ciriolo, A. G., Sciarrino, F., \& Osellame, R.}
\bibinfo{title} {Fast escape of a quantum walker from an integrated photonic maze.}
\newblock \emph{\bibinfo{journal}{Nat. Commun.}}
\textbf{\bibinfo{volume}{7}}, \bibinfo{pages}{11682}
  (\bibinfo{year}{2016}).

\bibitem{Tang2019}
\bibinfo{author}{Tang, H., Feng, Z., Wang, Y. H., Lai, P. C., Wang, C. Y., Ye, Z. Y., Wang, C. K., Shi, Z. Y., Wang, T. Y., Chen, Y., Gao, J. \& Jin, X., M.}
\newblock \bibinfo{title}{Experimental quantum stochastic walks simulating associative memory of Hopfield neural networks.}
\newblock \emph{\bibinfo{journal}{Phys. Rev. Applied}}
\textbf{\bibinfo{volume}{11}}, \bibinfo{pages}{024020}
(\bibinfo{year}{2019}).

\bibitem{Perez2018}
\bibinfo{author} {Perez-Leija, A., Guzm\'an-Silva, D., de J. Le\'on-Montiel, R., Gr\"afe, M., Heinrich, M., Moya-Cessa, H., Busch, K.,\& Szameit, A.} 
\bibinfo{title} {Endurance of quantum coherence due to particle indistinguishability in noisy quantum networks.}
\newblock \emph{\bibinfo{journal}{npj Quantum Info.}}
\textbf{\bibinfo{volume}{4}}, \bibinfo{pages}{45}
  (\bibinfo{year}{2018}).


\bibitem{Tang2022}
\bibinfo{author} {Tang, H., Banchi, L., Wang, T.Y., Shang, X.W., Tan, X., Zhou, W.H., Feng, Z., Pal, A., Li, H., Hu, C.Q., Kim, M.S. \& Jin, X. M.} 
\bibinfo{title} {Generating Haar-uniform randomness using stochastic quantum walks on a photonic chip.}
\newblock \emph{\bibinfo{journal}{Phys. Rev. Lett.}}
\textbf{\bibinfo{volume}{128}}, \bibinfo{pages}{050503}
  (\bibinfo{year}{2022}).
  
  

\bibitem{Tang2018}
\bibinfo{author}{Tang, H., Lin, X. F., Feng, Z., Chen, J. Y., Gao, J., Sun, K., Wang, C. Y., Lai, P. C., Xu, X. Y., Wang, Y., Qiao, L. F., Yang, A. L., \& Jin, X., M.}
\newblock \bibinfo{title}{Experimental Two-dimensional Quantum Walk on a Photonic Chip.}
\newblock \emph{\bibinfo{journal}{Sci. Adv.}}
\textbf{\bibinfo{volume}{4}}, \bibinfo{pages}{eaat3174}
(\bibinfo{year}{2018}).

\bibitem{Tang2018b}
\bibinfo{author}{Tang, H., Di Franco, C., Shi, Z. Y., He, T. S., Feng, Z., Gao, J., Li, Z. M., Jiao Z. Q., Wang, T. Y., Kim, M. S.,\& Jin, X. M.}
\newblock \bibinfo{title}{Experimental quantum fast hitting on hexagonal graphs.}
\newblock \emph{\bibinfo{journal}{Nat. Photon.}}
\textbf{\bibinfo{volume}{12}}, \bibinfo{pages}{754-758}
(\bibinfo{year}{2018}).

 \bibitem{Wang2022}
\bibinfo{author} {Wang, Y., Lu, Y.H., Gao, J., Chang, Y.J., Ren, R.J., Jiao, Z.Q., Zhang, Z.Y., \& Jin, X. M.} 
\bibinfo{title} {Topologically Protected Polarization Quantum Entanglement on a Photonic Chip.}
\newblock \emph{\bibinfo{journal}{Chip}}
\textbf{\bibinfo{volume}{1}}, \bibinfo{pages}{100003}
  (\bibinfo{year}{2022}).

\bibitem{PDB}
\newblock \bibinfo{title}{Image of 3BSD (Ben-Shem, A., Frolow, F. and Nelson, N. Evolution of photosystem I - from symmetry through pseudosymmetry to asymmetry. \emph{FEBS letters} \textbf{564}, 274-280 (2004)) created with VMD (Humphrey, W., Dalke, A. and Schulten, K., VMD - Visual Molecular Dynamics. \emph{J. Molec. Graphics} \textbf{14}, 33-38 (1996)).} 

\bibitem{Adolphs2006}
\bibinfo{author} {Adolphs, J. \& Renger, T.} 
\bibinfo{title} {How proteins trigger excitation energy transfer in the FMO complex of green sulfur bacteria.}
\newblock \emph{\bibinfo{journal}{Biophys. J.}}
\textbf{\bibinfo{volume}{91}}, \bibinfo{pages}{2778-2797}
  (\bibinfo{year}{2006}).

\bibitem{Hoyer2010}
\bibinfo{author} {Hoyer, S., Sarovar, M. \& Whaley, K.B.} 
\bibinfo{title} {Limits of quantum speedup in photosynthetic light harvesting.}
\newblock \emph{\bibinfo{journal}{New J. Phys.}}
\textbf{\bibinfo{volume}{12}}, \bibinfo{pages}{065041}
  (\bibinfo{year}{2010}).

  
\bibitem{Gelin2019}
\bibinfo{author} {Gelin, M. F., Borrelli, R., \& Domcke, W.} 
\bibinfo{title} {Origin of unexpectedly simple oscillatory responses in the excited-state dynamics of disordered molecular aggregates.}
\newblock \emph{\bibinfo{journal}{J. Phys. Chem. Lett.}}
\textbf{\bibinfo{volume}{10}}, \bibinfo{pages}{2806-2810}
  (\bibinfo{year}{2019}). 

\bibitem{WangL2019}
\bibinfo{author} {Wang, L., Allodi, M.A., \& Engel, G.S.} 
\bibinfo{title} {Quantum coherences reveal excited-state dynamics in biophysical systems.}
\newblock \emph{\bibinfo{journal}{Nat. Rev. Chem.}}
\textbf{\bibinfo{volume}{3}}, \bibinfo{pages}{477-490}
  (\bibinfo{year}{2019}). 

\bibitem{Coates2021}
\bibinfo{author} {Coates, A.R., Lovett, B.W., \& Gauger, E.M.} 
\bibinfo{title} {Localisation determines the optimal noise rate for quantum transport.}
\newblock \emph{\bibinfo{journal}{New J. Phys.}}
\textbf{\bibinfo{volume}{23}}, \bibinfo{pages}{123014}
  (\bibinfo{year}{2021}). 


\bibitem{Jeong2004}
\bibinfo{author} {Jeong, H., Paternostro, M., \& Kim, M. S.} 
\bibinfo{title} {Simulation of quantum random walks using interference of classical field.}
\newblock \emph{\bibinfo{journal}{Phys. Rev. A}}
\textbf{\bibinfo{volume}{69}}, \bibinfo{pages}{012310}
  (\bibinfo{year}{2004}).



\bibitem{Novoderezhkin2010}
\bibinfo{author} {Novoderezhkin, V. I., Doust, A. B., Curutchet, C., Schole, G. D., \& van Grondelle, R.}
\bibinfo{title} {Excitation Dynamics in Phycoerythrin 545: Modeling of Steady-State Spectra and Transient Absorption with Modified Redfield Theory.}
\newblock \emph{\bibinfo{journal}{Biophysical Journal}}
\textbf{\bibinfo{volume}{99}}, \bibinfo{pages}{344-352}
(\bibinfo{year}{2010})

\bibitem{Chandrasekaran2016}
\bibinfo{author} {Chandrasekaran, S., Pothula, K. R., \& Kleinekathöfer, U.}
\bibinfo{title} {Protein Arrangement Effects on the Exciton Dynamics in the PE555 Complex.}
\newblock \emph{\bibinfo{journal}{Journal of Physical Chemistry B}}
\textbf{\bibinfo{volume}{121}}, \bibinfo{pages}{3228}
(\bibinfo{year}{2016})

\bibitem{Zech2014}
\bibinfo{author} {Zech, T., Mulet, R., Wellens, T., \& Buchleitner, A.}
\bibinfo{title} {Centrosymmetry enhances quantum transport in disordered molecular networks.}
\newblock \emph{\bibinfo{journal}{New J. Phys.}}
\textbf{\bibinfo{volume}{16}}, \bibinfo{pages}{055002}
(\bibinfo{year}{2014})

\bibitem{Choi2021}
\bibinfo{author} {Choi, E. H., Uhm, H. S., \& Kaushik, N. K.} 
\bibinfo{title} {Plasma bioscience and its application to medicine.}
\newblock \emph{\bibinfo{journal}{AAPPS Bulletin}}
\textbf{\bibinfo{volume}{31}}, \bibinfo{pages}{10}
  (\bibinfo{year}{2021}).
  

\bibitem{Hu2022}
\bibinfo{author} {Hu, Z. X., Head-Marsden, K., Mazziotti, D. A., Narang, P., \& Kais, S.} 
\bibinfo{title} {A general quantum algorithm for open quantum dynamics demonstrated with the Fenna-Matthews-Olson complex.}
\newblock \emph{\bibinfo{journal}{Quantum}}
\textbf{\bibinfo{volume}{6}}, \bibinfo{pages}{726}
  (\bibinfo{year}{2022}). 
  
  
\bibitem{Georgescu2014}
\bibinfo{author}{Georgescu, I. M., Ashhab, S., \& Nori, F.}
\newblock \bibinfo{title}{Quantum simulation.}
\newblock \emph{\bibinfo{journal}{Rev. Mod. Phys.}}
  \textbf{\bibinfo{volume}{86}}, \bibinfo{pages}{153-185} (\bibinfo{year}{2014}).


\bibitem{Chen2018}
\bibinfo{author} {Chen, Y., Gao, J., Jiao, Z. Q., Sun, K., Qiao, L. F., Tang, H., Lin, X. F., \& Jin, X. M.} 
\bibinfo{title} {Mapping Twisted Light into and out of a Photonic Chip.}
\newblock \emph{\bibinfo{journal}{Phys. Rev. Lett.}}
\textbf{\bibinfo{volume}{121}}, \bibinfo{pages}{233602}
  (\bibinfo{year}{2018}).
 
\bibitem{Wang2019b}
\bibinfo{author} {Wang, Y., Gao, J., Pang, X. L., Jiao, Z. Q., Tang, H., Chen, Y., Qiao, L. F., Gao, Z. W., Dou, J. P., Yang, A. L., \& Jin, X. M.} 
\bibinfo{title} {Experimental Parity-Induced Thermalization Gap in Disordered Ring Lattices.}
\newblock \emph{\bibinfo{journal}{Phys. Rev. Lett.}}
\textbf{\bibinfo{volume}{122}}, \bibinfo{pages}{013903}
  (\bibinfo{year}{2019}).

\bibitem{Wang2020}
\bibinfo{author} {Wang, Y., Sheng, C., Lu, Y.H., Gao, J., Chang, Y.J., Pang, X.L., Yang, T.H., Zhu, S.N., Liu, H. \& Jin, X.M.} 
\bibinfo{title} {Quantum simulation of particle pair creation near the event horizon.}
\newblock \emph{\bibinfo{journal}{National Science Review}}
\textbf{\bibinfo{volume}{7}}, \bibinfo{pages}{1476-1484}
  (\bibinfo{year}{2020}).

 
\bibitem{Tang2022b}
\bibinfo{author} {Tang, H., Wang, T.Y., Shi, Z.Y., Feng, Z., Wang, Y., Shang, X.W., Gao, J., Jiao, Z.Q., Li, Z.M., Chang, Y.J. and Zhou, W.H., Lu, Y. H., Yang, Y. L., Ren, R. J., Qiao, L. F., \& Jin, X. M.} 
\bibinfo{title} {Experimental quantum simulation of dynamic localization on curved photonic lattices.}
\newblock \emph{\bibinfo{journal}{Photonics Research}}
\textbf{\bibinfo{volume}{10}}, \bibinfo{pages}{1430-1439}
  (\bibinfo{year}{2022}).
  
\bibitem{Pirandola2021}
\bibinfo{author} {Pirandola, S., Ottaviani, C., Jacobsen, C. S., Spedalieri, G., Braunstein, S. L., Gehring, T., \& Andersen, U. L.} 
\bibinfo{title} {Environment-assisted bosonic quantum communications.}
\newblock \emph{\bibinfo{journal}{npj Quantum Information}}
\textbf{\bibinfo{volume}{7}}, \bibinfo{pages}{1-7}
  (\bibinfo{year}{2021}).

\bibitem{Andersson2019}
\bibinfo{author} {Andersson, G., Suri, B., Guo, L., Aref, T., \& Delsing, P.} 
\bibinfo{title} {Non-exponential decay of a giant artificial atom.}
\newblock \emph{\bibinfo{journal}{Nat. Phys.}}
\textbf{\bibinfo{volume}{15}}, \bibinfo{pages}{1123-1127}
  (\bibinfo{year}{2019}).







\end{thebibliography}

\begin{thebibliography}{10}
\renewcommand{\bibnumfmt}[1]{[S#1]}
\bibitem{Adolphs2006}
\bibinfo{author} {Adolphs, J. \& Renger, T.} 
\bibinfo{title} {How proteins trigger excitation energy transfer in the FMO complex of green sulfur bacteria.}
\newblock \emph{\bibinfo{journal}{Biophys. J.}}
\textbf{\bibinfo{volume}{91}}, \bibinfo{pages}{2778-2797}
  (\bibinfo{year}{2006}).

\bibitem{Hoyer2010}
\bibinfo{author} {Hoyer, S., Sarovar, M. \& Whaley, K.B.} 
\bibinfo{title} {Limits of quantum speedup in photosynthetic light harvesting.}
\newblock \emph{\bibinfo{journal}{New J. Phys.}}
\textbf{\bibinfo{volume}{12}}, \bibinfo{pages}{065041}
  (\bibinfo{year}{2010}).
  
\bibitem{Hoyer2016}
\bibinfo{title} {A personal conversation with S. Hoyer in 2016 inquiring the sign of the coupling coefficients for FMO complex.}

\bibitem{Novoderezhkin2010}
\bibinfo{author} {Novoderezhkin, V. I., Doust, A. B., Curutchet, C., Schole, G. D., \& van Grondelle, R.}
\bibinfo{title} {Excitation Dynamics in Phycoerythrin 545: Modeling of Steady-State Spectra and Transient Absorption with Modified Redfield Theory.}
\newblock \emph{\bibinfo{journal}{Biophysical Journal}}
\textbf{\bibinfo{volume}{99}}, \bibinfo{pages}{344-352}
(\bibinfo{year}{2010})

\bibitem{Chandrasekaran2016}
\bibinfo{author} {Chandrasekaran, S., Pothula, K. R., \& Kleinekathöfer, U.}
\bibinfo{title} {Protein Arrangement Effects on the Exciton Dynamics in the PE555 Complex.}
\newblock \emph{\bibinfo{journal}{Journal of Physical Chemistry B}}
\textbf{\bibinfo{volume}{121}}, \bibinfo{pages}{3228}
(\bibinfo{year}{2016})

\bibitem{Zech2014}
\bibinfo{author} {Zech, T., Mulet, R., Wellens, T., \& Buchleitner, A.}
\bibinfo{title} {Centrosymmetry enhances quantum transport in disordered molecular networks.}
\newblock \emph{\bibinfo{journal}{New J. Phys.}}
\textbf{\bibinfo{volume}{16}}, \bibinfo{pages}{055002}
(\bibinfo{year}{2014})


\bibitem{Crespi2013}
\bibinfo{author} {Crespi, A., Osellame, R., Ramponi, R., Brod, D. J., Galv\~ao, E. F., Spagnolo, N., Vitelli, C., Maiorino, E., Mataloni, P., \& Sciarrino, F.}
\bibinfo{title} {Integrated multimode interferometers with arbitrary designs for photonic boson sampling.}
\newblock \emph{\bibinfo{journal}{Nat. Photon.}}
\textbf{\bibinfo{volume}{7}}, \bibinfo{pages}{545-549}
(\bibinfo{year}{2013})

\bibitem{Chaboyer2015}
\bibinfo{author} {Chaboyer, Z., Meany, T., Helt, L. G., Withford, M. J., \& Steel, M. J.}
\bibinfo{title} {Tunable quantum interference in a 3D integrated circuit.}
\newblock \emph{\bibinfo{journal}{Sci. Rep.}}
\textbf{\bibinfo{volume}{5}}, \bibinfo{pages}{9601}
(\bibinfo{year}{2015})

\bibitem{Tang2018}
\bibinfo{author} {Tang, H., Lin, X. F., Feng, Z., Chen, J. Y., Gao, J., Sun, K., Wang, C. Y., Lai, P. C., Xu, X. Y., Wang, Y., Qiao, L. F., Yang, A. L., \& Jin, X., M.}
\bibinfo{title} {Experimental Two-dimensional Quantum Walk on a Photonic Chip.}
\newblock \emph{\bibinfo{journal}{Sci. Adv.}}
\textbf{\bibinfo{volume}{4}}, \bibinfo{pages}{eaat3174}
(\bibinfo{year}{2018})

\bibitem{Szameit2007}
\bibinfo{author}{Szameit, A., Dreisow, F., Pertsch, T., Nolte, S.,\& Trnnermann, A. }
\newblock \bibinfo{title}{Control of directional evanescent coupling in fs laser written waveguides.}
\newblock \emph{\bibinfo{journal}{Opt. Express}}
\textbf{\bibinfo{volume}{15}}, \bibinfo{pages}{1579-1587}
(\bibinfo{year}{2007}).

\bibitem{Lebugle2015}
\bibinfo{author}{ Lebugle, M., Gr\"afe, M., Heilmann, R., Perez-Leija, A., Nolte, S., \& Szameit, A. }
\newblock \bibinfo{title}{Experimental observation of n00n state bloch oscillations.}
\newblock \emph{\bibinfo{journal}{Nat. Commun.}}
\textbf{\bibinfo{volume}{6}}, \bibinfo{pages}{8273}
(\bibinfo{year}{2015}).



\bibitem{Box2008}
\bibinfo{author} {George, E., P., Box, Gwilym, M., Jenkins, \& Gregory, C., Reinsel} 
\bibinfo{title} {Time Series Analysis : Forecasting and Control (4th ed.).}
\newblock \emph{\bibinfo{book}{John Wiley \& Sons, New Jersey.}}
\textbf{\bibinfo{volume}{1}}, \bibinfo{pages}{31-33}
(\bibinfo{year}{2008}).

\bibitem{Markus2000}
\bibinfo{author} {Markus, W., T$\tilde{o}$nu, P., Milosz A. P., Simone I. E.V., Thijs J. A., Rienk v. G., \& Herbert v. A.} 
\bibinfo{title} {Electron-Vibrational Coupling in the Fenna-Matthews-Olson Complex of Prosthecochloris aestuarii Determined by Temperature-Dependent Absorption and Fluorescence Line-Narrowing Measurements.}
\newblock \emph{\bibinfo{journal}{J. Phys. Chem. B}}
\textbf{\bibinfo{volume}{104}}, \bibinfo{pages}{5825-5831}
  (\bibinfo{year}{2000}).
  
\bibitem{Alexander2020}
\bibinfo{author} {Alexander, K., Dominik, L., Frankm, M., \& Thomas, R.} 
\bibinfo{title}{Normal mode analysis of spectral density of FMO trimers: Intra- and intermonomer energy transfer.}
\newblock \emph{\bibinfo{journal}{J. Chem. Phys.}}
\textbf{\bibinfo{volume}{153}}, \bibinfo{pages}{215103}
  (\bibinfo{year}{2020}).

\bibitem{Cao2020}
\bibinfo{author} {Cao, J. S., Cogdell, R. J., Coker, D. F., Duan, H. G., Hauer, J., Kleinekath$\ddot{o}$fer, U., Jansen, T. L. C., Man$\check{c}$al, T., Miller, R. J. D., Ogilvie, J. P., Prokhorenko, V. I., Renger, T., Tan, H. S., Temoelaar, R., Thorwart, M., Thyrhaug, E., Westenhoff, S., \& Zigmantas, D.} 
\bibinfo{title} {Quantum biology revisited.}
\newblock \emph{\bibinfo{journal}{Sci. Adva.}}
\textbf{\bibinfo{volume}{6}}, \bibinfo{pages}{eaaz4888}
  (\bibinfo{year}{2020}).

\bibitem{Forster1948}
\bibinfo{author} {F{\"o}rster, T.} 
\bibinfo{title} {Zwischenmolekulare Energiewanderung und Fluoreszenz.}
\newblock \emph{\bibinfo{journal}{Annalen der Physik}}
\textbf{\bibinfo{volume}{437}}, \bibinfo{pages}{55-75}
  (\bibinfo{year}{1948}).

\bibitem{Stryer1967}
\bibinfo{author} {Stryer, L. \& Haugland, R. P.} 
\bibinfo{title} {Energy transfer: a spectroscopic ruler.}
\newblock \emph{\bibinfo{journal}{Proceedings of the National Academy of Sciences of the United States of America}}
\textbf{\bibinfo{volume}{58}}, \bibinfo{pages}{719-726}
  (\bibinfo{year}{1967}).

\bibitem{Park2016}
\bibinfo{author} {Park, H., Heldman, N., Rebentrost, P., Abbondanza, L., Iagatti, A., Alessi, A., Patrizi, B., Salvalaggio, M., Bussotti, L., Mohseni, M., Caruso, F., Johnsen, H. C., Fusco, R., Foggi, P., Scudo, P. F., Lloyd, S., \& Belcher, A. M.} 
\bibinfo{title} {Enhanced energy transport in genetically engineered excitonic networks.}
\newblock \emph{\bibinfo{journal}{Nature Materials}}
\textbf{\bibinfo{volume}{15}}, \bibinfo{pages}{211-216}
  (\bibinfo{year}{2016}).


\bibitem{Plenio2008}
\bibinfo{author} {Plenio, M. B. \& Huelga, S. F.} 
\bibinfo{title} {Dephasing-assisted transport: quantum networks and biomolecules.}
\newblock \emph{\bibinfo{journal}{New Journal of Physics}}
\textbf{\bibinfo{volume}{10}}, \bibinfo{pages}{113019}
  (\bibinfo{year}{2008}).


\bibitem{Mohseni2008}
\bibinfo{author} {Mohseni, M., Rebentrost, P., Lloyd, S., \& Aspuru-Guzik, A.} 
\bibinfo{title} {Environment-assisted quantum walks in photosynthetic energy transfer.}
\newblock \emph{\bibinfo{journal}{The Journal of Chemical Physics}}
\textbf{\bibinfo{volume}{129}}, \bibinfo{pages}{174106}
  (\bibinfo{year}{2008}).
  

\bibitem{Lambert2013}
\bibinfo{author} {Lambert, N., Chen, Y.-N., Cheng, Y.-C., Li, C.-M., Chen, G.-Y., \& Nori, F.} 
\bibinfo{title} {Quantum biology.}
\newblock \emph{\bibinfo{journal}{Nature Physics}}
\textbf{\bibinfo{volume}{9}}, \bibinfo{pages}{10-18}
  (\bibinfo{year}{2013}).


\bibitem{Kassal2013}
\bibinfo{author} {Kassal, I., Yuen-Zhou, J., \& Rahimi-Keshari, S.} 
\bibinfo{title} {Does Coherence Enhance Transport in Photosynthesis?}
\newblock \emph{\bibinfo{journal}{The Journal of Physical Chemistry Letters}}
\textbf{\bibinfo{volume}{4}}, \bibinfo{pages}{362-367}
  (\bibinfo{year}{2013}).


\bibitem{Engel2007}
\bibinfo{author} {Engel, G. S., Calhoun, T. R., Read, E. L., Ahn, T.-K., Man$\check c$al, T., Cheng, Y.-C., Blankenship, R. E., \& Fleming, G. R.} 
\bibinfo{title} {Evidence for wavelike energy transfer through quantum coherence in photosynthetic systems.}
\newblock \emph{\bibinfo{journal}{Nature}}
\textbf{\bibinfo{volume}{446}}, \bibinfo{pages}{782-786}
  (\bibinfo{year}{}).
  
  
\bibitem{Chin2010}
\bibinfo{author} {Chin, A. W., Datta, A., Caruso, F., Huelga, S. F., \& Plenio, M. B.} 
\bibinfo{title} {Noise-assisted energy transfer in quantum networks and light-harvesting complexes.}
\newblock \emph{\bibinfo{journal}{New Journal of Physics}}
\textbf{\bibinfo{volume}{12}}, \bibinfo{pages}{065002}
  (\bibinfo{year}{2010}).


\bibitem{Harush2021}
\bibinfo{author} {Harush, E. Z. \& Dubi, Y.} 
\bibinfo{title} {Do photosynthetic complexes use quantum coherence to increase their efficiency? Probably not.}
\newblock \emph{\bibinfo{journal}{Science Advances}}
\textbf{\bibinfo{volume}{7}}, \bibinfo{pages}{eabc4631}
  (\bibinfo{year}{2021}).


\bibitem{Stones2016}
\bibinfo{author} {Stones, R. \& Olaya-Castro, A.} 
\bibinfo{title} {Vibronic Coupling as a Design Principle to Optimize Photosynthetic Energy Transfer.}
\newblock \emph{\bibinfo{journal}{Chem}}
\textbf{\bibinfo{volume}{1}}, \bibinfo{pages}{822-824}
  (\bibinfo{year}{2016}).


\bibitem{Higgins2021}
\bibinfo{author} {Higgins, J. S., Lloyd, L. T., Sohail, S. H., Allodi, M. A., Otto, J. P., Saer, R. G., Wood, R. E., Massey, S. C., Ting, P.-C., Blankenship, R. E., \& Engel, G. S.} 
\bibinfo{title} {Photosynthesis tunes quantum-mechanical mixing of electronic and vibrational states to steer exciton energy transfer.}
\newblock \emph{\bibinfo{journal}{Proceedings of the National Academy of Sciences}}
\textbf{\bibinfo{volume}{118}}, \bibinfo{pages}{e2018240118}
  (\bibinfo{year}{2021}).


\bibitem{Duan2017}
\bibinfo{author} {Duan, H.-G., Prokhorenko, V. I., Cogdell, R. J., Ashraf, K., Stevens, A. L., Thorwart, M., \& Miller, R. J. D.} 
\bibinfo{title} {Nature does not rely on long-lived electronic quantum coherence for photosynthetic energy transfer.}
\newblock \emph{\bibinfo{journal}{Proceedings of the National Academy of Sciences}}
\textbf{\bibinfo{volume}{114}}, \bibinfo{pages}{8493-8498}
  (\bibinfo{year}{2017}).





  
  
\bibitem{Rebentrost2009}
\bibinfo{author} {Rebentrost, P., Mohseni, M., Kassal, I., Lloyd, S., \& Aspuru-Guzik, A.} 
\bibinfo{title} {Environment-assisted quantum transport.}
\newblock \emph{\bibinfo{journal}{New Journal of Physics}}
\textbf{\bibinfo{volume}{11}}, \bibinfo{pages}{033003}
  (\bibinfo{year}{2009}).

\bibitem{Jeong2004}
\bibinfo{author} {Jeong, H., Paternostro, M., \& Kim, M. S.} 
\bibinfo{title} {Simulation of quantum random walks using interference of classical field.}
\newblock \emph{\bibinfo{journal}{Phys. Rev. A}}
\textbf{\bibinfo{volume}{69}}, \bibinfo{pages}{012310}
  (\bibinfo{year}{2004}).


\bibitem{Anderson1958}
\bibinfo{author} {Anderson, P. W.} 
\bibinfo{title} {Absence of Diffusion in Certain Random Lattices.}
\newblock \emph{\bibinfo{journal}{Physical Review}}
\textbf{\bibinfo{volume}{109}}, \bibinfo{pages}{1492-1505}
  (\bibinfo{year}{1958}).
  

\bibitem{Coates2021}
\bibinfo{author} {Coates, A. R., Lovett, B. W., \& Gauger, E. M.} 
\bibinfo{title} {Localisation determines the optimal noise rate for quantum transport.}
\newblock \emph{\bibinfo{journal}{New Journal of Physics}}
\textbf{\bibinfo{volume}{23}}, \bibinfo{pages}{123014}
  (\bibinfo{year}{2021}).


\bibitem{Caruso2009}
\bibinfo{author} {Caruso, F., Chin, A. W., Datta, A., Huelga, S. F., \& Plenio, M. B.} 
\bibinfo{title} {Highly efficient energy excitation transfer in light-harvesting complexes: The fundamental role of noise-assisted transport.}
\newblock \emph{\bibinfo{journal}{The Journal of Chemical Physics}}
\textbf{\bibinfo{volume}{131}}, \bibinfo{pages}{105106}
  (\bibinfo{year}{2009}).
  

\bibitem{Zerah2020}
\bibinfo{author} {Zerah-Harush, E. \& Dubi, Y.} 
\bibinfo{title} {Effects of disorder and interactions in environment assisted quantum transport.}
\newblock \emph{\bibinfo{journal}{Physical Review Research}}
\textbf{\bibinfo{volume}{2}}, \bibinfo{pages}{023294}
  (\bibinfo{year}{2020}).
  
  

  
\bibitem{Tang2022}
\bibinfo{author} {Tang, H., Banchi, L., Wang, T.Y., Shang, X.W., Tan, X., Zhou, W.H., Feng, Z., Pal, A., Li, H., Hu, C.Q., Kim, M.S., \& Jin, X. M.} 
\bibinfo{title} {Generating Haar-uniform randomness using stochastic quantum walks on a photonic chip.}
\newblock \emph{\bibinfo{journal}{Phys. Rev. Lett.}}
\textbf{\bibinfo{volume}{128}}, \bibinfo{pages}{050503}
  (\bibinfo{year}{2022}).
  
  







\end{thebibliography}
\end{document}